\documentclass[published]{JHEP3}
\JHEP{09(2003)027}

\usepackage{epsfig,bbm}

\newcommand{\SU}{\mathop{\rm SU}\nolimits}
\newcommand{\UU}{\mathop{\rm {}U}\nolimits}
\newcommand{\Tr}{\mathop{\rm Tr}\nolimits}

\newcommand{\cZ}{{\cal Z}}
\newcommand{\cF}{{\cal F}}
\newcommand{\cG}{{\cal G}}
\newcommand{\Z}{{\mathbb Z}}
\newcommand{\eq}[1]{(\ref{#1})}
\newcommand{\dd}{\mbox{d}}
\newcommand{\dD}{{\cal D}}
\def\bbbone{\mathbbm{1}}

\title{Blocking of lattice monopoles from the continuum
in hot lattice gluodynamics}

\author{Maxim~N.~Chernodub,$^{ab}$ Katsuya~Ishiguro$^b$ and Tsuneo~Suzuki$^b$\\
        \llap{$^a$}Institute of Theoretical and  Experimental Physics\\
	B.~Cheremushkinskaja 25, Moscow, 117259, Russia\\
	\llap{$^b$}Institute for Theoretical Physics, Kanazawa University\\
	Kanazawa 920-1192, Japan\\
	E-mail: \email{maxim@heron.itep.ru},
	        \email{ishiguro@hep.s.kanazawa-u.ac.jp},
	        \email{suzuki@hep.s.kanazawa-u.ac.jp}}

\abstract{The abelian monopoles in lattice gluodynamics are associated
  with continuum monopoles blocked to the lattice. This association
  allows to predict the lattice monopole action and density of the
  (squared) monopole charges from a continuum monopole model. The
  method is applied to the static monopoles in high temperature
  gluodynamics. We show that the numerical data both for the density
  and the action of the lattice monopoles can be described in terms of
  a Coulomb gas of abelian monopoles in the continuum.}

\received{June 6, 2003}
\revised{August 28, 2003}
\accepted{September 10, 2003}

\begin{document}

\section{Introduction}

One of the most popular approaches to the problem of confinement of
quarks in QCD is the so-called dual superconductor
mechanism~\cite{DualSuperconductor}. The key role in this approach is
played by abelian monopoles which are identified with the help of the
abelian projection method~\cite{AbelianProjections}. The basic idea
behind the abelian projections is to fix partially the non-abelian
gauge symmetry up to an abelian subgroup. For $\SU(N)$ gauge theories
the residual abelian symmetry group is compact since the original
non-abelian group is compact as well. The abelian monopoles arise
naturally due to the compactness of the residual gauge subgroup.

In the dual superconductor mechanism the abelian monopoles are
considered as effective degrees of freedom which are responsible for
confinement of quarks. According to the numerical
results~\cite{MonopoleCondensation} the monopoles are condensed in the
low temperature (confinement) phase. The condensation of the monopoles
leads to formation of the chromoelectric string due to the dual
Meissner effect. As a result the fundamental sources of chromoelectric
field, quarks, are confined by the string. The importance of the
abelian monopoles is also stressed by the
existence~\cite{AbelianDominance} of the abelian dominance phenomena
which was first observed in the lattice $\SU(2)$ gluodynamics: the
monopoles in the so-called Maximal abelian projection~\cite{MaA} make
a dominant contribution to the zero temperature string tension (for a
review, see ref.~\cite{Review}).

In the deconfinement phase (high temperatures) the monopoles are
not condensed and the quarks are liberated. This does not mean,
however, that monopoles do not play a role in non-perturbative
physics. It is known that in the deconfinement phase the vacuum is
dominated by static monopoles (which run along the ``temperature"
direction in the euclidean theory) while monopoles running in
spatial directions are suppressed. The static monopoles should
contribute to the ``spatial string tension" --- a coefficient in
front of the area term of large spatial Wilson loops. And,
according to numerical simulations in the deconfinement
phase~\cite{AbelianDominanceT}, the monopoles make a dominant
contribution to the spatial string tension.\footnote{Note that in another approach
to the ``spatial confinement" problem, the area law for the spatial Wilson loops
is suggested to be caused by magnetic thermal quasi-particles discussed in
ref.~\cite{Giovannangeli:2001bh}. It seems plausible that the continuum abelian
monopoles may be correlated with these magnetic excitations.}

In this paper we investigate the physics of the lattice monopole
currents in the $\SU(2)$ gauge theory at high temperatures. The
properties of the monopoles are effectively three dimensional due to
static nature of the currents. However, the problem which is discussed
in this article is quite general and it can not be ascribed only to
the three dimensional case which is considered here as a useful and
physically relevant example.

The lattice monopoles in Monte Carlo simulations are detected by the
standard De\-Grand-Tous\-saint construction~\cite{DGT} which
identifies the magnetic monopoles by measuring the magnetic flux
coming out of lattice $3D$ cells (cubes). The properties of
thus-measured lattice monopoles should obviously depend on the
physical size, $b$, of the lattice cell (below we call these lattice
objects as ``lattice monopoles of the size $b$").  To get the continuum
properties of the monopoles one should send the size of the lattice
cells to zero, $b \to 0$, what is usually a difficult numerical
problem. Here we propose a method --- which can be called as ``blocking
of monopoles from continuum to the lattice" --- to identify the
couplings of a continuum model for monopoles using the numerical
results obtained on the lattice with a finite spacing.

Each lattice $3D$ cell can be considered as a ``detector" which
measures the magnetic charges of the continuum monopoles travelling
through the lattice: if the continuum monopole is located inside a
lattice $3D$ cell then the DeGrand-Toussaint method detects a
\emph{lattice} monopole. From this point of view the limit $b \to 0$
is needed to detect accurately the position of the continuum monopole
with the help of the lattice detector. If the size of the lattice cell
is finite then two or more continuum monopoles may be placed inside
the cell. The fluctuations of the monopole charges of the lattice
cells should be described by a continuum model. As a result, the
lattice observables --- such as the vacuum expectation value of the
lattice monopole density --- should carry information about dynamics
of the continuum monopoles. The observables should depend not only on
the size of the lattice cell, $b$, but also on features of the
continuum model which describes the monopole dynamics. We take the
high temperature gluodynamics as a simple example and show that the
density and action of the static lattice monopoles of various sizes
$b$ can self-consistently be described by a continuum Coulomb gas
model.

Our approach resembles the idea of the blocking of the continuum
\emph{fields} to the
lattice~\cite{BlockingOfFields}--\cite{BlockingOfTopology}.  This
method allows to construct perfect actions and operators in various
field theories. In our paper we are blocking continuum monopoles what
is ideologically similar to the blocking of a topological
charge~\cite{BlockingOfTopology} and a fermionic
current~\cite{BlockingOfFields,Wolfgang2} to the lattice. The blocking
of the fields, however, leads to a \emph{non-integer} lattice magnetic
current contrary to the quantized lattice definition of the monopole
charge~\cite{DGT}. Thus the blocking of topological defects seems to
be more suitable for investigation of the lattice monopole
charges. There is also a similarity of our approach with the blocking
of the lattice monopole degrees of freedom from fine to coarser
lattices~\cite{BlockSpin}. This method allows to define a perfect
monopole action independent on the spacing of the fine lattice.

To avoid misunderstanding we would like to stress from the very
beginning the difference between various lattice monopole sizes $b$.
As we mentioned above, we call the size of the lattice $3D$ cell ---
used for detection of the monopoles --- as ``the size of the lattice
monopole". This should be distinguished from the \emph{physical}
radius, $r_0$, of the monopole core~\cite{FiniteRadius} which is
obviously independent on the size of the lattice ``detector". In this
paper we disregard the existence of the monopole core and consider the
continuum monopoles as point-like objects. Thus, we get the
finite-sized lattice monopoles by blocking the point-like monopoles
from continuum to the lattice.

There is another motivation for our study. Various properties of the
lattice monopoles such as the action and density of the monopole
currents, depend on the \emph{physical} size of the lattice
monopole~\cite{DependenceOnLatticeSize,NumericalMonopoleAction}.  This
effect is observed both at zero~\cite{DependenceOnLatticeSize} and
finite~\cite{NumericalMonopoleAction} temperatures. Usually the
monopole density is calculated numerically for so called elementary
monopoles the size of which coincide with the lattice spacing
$a$. Then the monopole quantities in continuum limit ($e.g.$,
continuum monopole density) are usually associated with the limit $a
\to 0$. However, since the monopole in the gluodynamics has a core of
the finite size~\cite{FiniteRadius} the continuum monopole may simply
``drop" through the lattice cells if $a \ll r_0$. Moreover, the scale
of the ultraviolet lattice artifacts coincides with the lattice UV
cut-off $a$, which, in turn, equal to the size of the elementary
monopoles. Therefore the lattice elementary may be affected by the
lattice UV artifacts. Both these reasons imply that (at least,
naively) there is a possibility that the continuum monopole density
may not correctly be calculated as the \emph{elementary} lattice
monopole density in the limit $a \to 0$. One of advantages of our
approach is that it allows to relate the \emph{lattice} monopole
quantities to the corresponding \emph{physical} quantities in
continuum limit at the \emph{finite} lattice spacing making irrelevant
the problems noticed above.

The structure of the paper is as follows. In section~\ref{theory} we
derive the lattice monopole action and the density of squared magnetic
charges. The basic assumption behind the derivation is that the
dynamics of the continuum monopoles --- which are blocked to the
lattice --- is described by the Coulomb gas model. In
section~\ref{experiment} we present the numerical results both for the
monopole action and for the density of squared magnetic charges. We
show that these quantities are in a good agreement with the
predictions of section~\ref{theory}. The comparison of the analytical
and numerical results allows us to calculate the product of the
abelian magnetic screening mass and the monopole density in the
continuum model. At the end of section~\ref{experiment} we check the
self-consistency of our results as well as validity of the Coulomb gas
description for the static monopole currents. Our conclusion is
presented in the last section~\ref{sec4}.

\section{Lattice monopoles from continuum monopoles}
\label{theory}

In this section we consider the blocking of the continuum monopoles to
the lattice in three space-time dimensions. Let us consider a lattice
with a finite lattice spacing $b$ which is embedded in the continuum
space-time. The cells of the lattice are defined as follows:
\begin{equation}
C_s = \left\{b \Bigl(s_i - \frac{1}{2}\Bigr) \leq x_i \leq
b \Bigl(s_i + \frac{1}{2}\Bigr)\,,\quad i=1,2,3 \right\},
\end{equation}
where $s_i$ is the lattice dimensionless coordinate and $x_i$
corresponds to the continuum coordinate.

Each lattice cell, $C_s$, detects the total magnetic charge, $k_s$, of
``continuum" monopoles inside it:
\begin{equation}
k_s = \int\limits_{C_s}\dd^3 x\, \rho(x)\,, \qquad
\rho(x) = \sum_a  q_a \, \delta^{(a)} (x - x^{(a)})\,,
\label{ks}
\end{equation}
where $\rho(x)$ is the density of the continuum monopoles, $x_a$ and
$q_a$ is the position and the charge (in units of a fundamental
magnetic charge, $g_M$) of $a^{\mathrm{th}}$ continuum monopole. In
three dimensions the monopoles are instanton-like objects and the
monopole trajectories have zero dimensionality (points). It is worth
stressing the difference between continuum and lattice monopoles: the
continuum monopoles are fundamental point-like objects in this
approach while the lattice monopoles correspond to the lattice cells
with non-zero total magnetic charges of continuum monopoles located
inside appropriate cells.

According to definitions~\eq{ks}, the lattice monopole charge shares
similar properties to the continuum monopole charge. The lattice
monopole charge $k_s$ is quantized, $k_s \in \Z$, and conserved in the
three-dimensional sense:
\begin{equation}
\sum\limits_{s \in \Lambda} k_s \equiv \int\limits_V \dd^3 x \,
\rho(x) = 0\,,
\end{equation}
if the continuum charge is conserved. Here $\Lambda$ and $V$ denote
the lattice and continuum volume occupied by the lattice,
respectively. In other words, the total magnetic charge of the lattice
monopole configuration is zero on a finite lattice with periodic
boundary which is considered in this paper.

Suppose that the dynamics of the continuum mo\-no\-po\-les is governed
by the $3D$ Cou\-lomb gas model:
\begin{equation}
\cZ = \sum\limits_{N=0}^\infty \frac{\zeta^N}{N!}
\Biggl[\prod\limits^N_{a=1} \int \dd^3 x^{(a)} \sum\limits_{q_a = \pm 1}\Biggr]
\exp\left\{ - \frac{g^2_M}{2} \sum_{a,b=1 \atop a \neq b}^N
q_a q_b \, D\left(x^{(a)}-x^{(b)}\right)\right\}.
\label{CoulombModel}
\label{Z1}
\end{equation}
The Coulomb interaction in eq.~\eq{Z1} is represented by the inverse
laplacian $D$: $ - \partial^2_i D(x) = \delta^{(3)}(x)$.  The density
of the continuum monopole is assumed to be low. The continuum monopole
charges therefore are restricted by the condition $|q_a| \leq 1$ which
means that these monopoles do not overlap. The average continuum
monopole density $\rho$ is controlled by the fugacity parameter
$\zeta$.

Note that the model~\eq{CoulombModel} does not exist without properly
defined ultraviolet cut-off. Indeed, the self-energy of the point-like
monopoles is a linearly divergent function due to the infinite nature
of the Coulomb interaction.  This divergence --- which is not
explicitly present in eq.~\eq{CoulombModel} --- should appear
manifestly in a field representation of the Coulomb gas model which we
need below (see, for example, the sin-Gordon representation for the
lattice monopole action, eq.~\eq{Z3}). The renormalized fugacity is,
$\zeta_{\mathrm{ren}} = \zeta\cdot \exp\{ g^2_m \slash (8 \pi \,
r_0)\}$, where the $r_0$ is the ultraviolet cut-off.  In our case this
cut-off is given by the size of the monopole core which is of the
order of 0.05~fm at zero temperature~\cite{FiniteRadius}. For
simplicity we omit the subscript ``ren" in the renormalized fugacity
below.

The magnetic charges in the Coulomb gas~\eq{Z1} are screened: at large
distances the two-point charge correlation function is exponentially
suppressed, $\langle \rho(x) \rho(y)\rangle \sim \exp\{- |x-y| \slash
\lambda_D\}$. Here $\lambda_D$ is the Debye screening
length~\cite{Polyakov},
\begin{equation}
\lambda_D = \frac{1}{g_M \sqrt{\rho}}\,.
\label{lambdaD}
\end{equation}
Note that the three dimensional Debye screening length corresponds to
a \emph{magnetic} screening in four dimensions. The density of the
continuum monopoles in the leading order is related to fugacity
as~\cite{Polyakov} $\rho = 2 \zeta$.  Below we choose the vacuum
expectation value of the continuum monopole density, $\rho$, and the
Debye screening length, $\lambda_D$, as suitable parameters of the
continuum model (instead of $g_M$ and $\zeta$).

We are interested in two basic quantities characterizing the lattice
mo\-no\-po\-les: the lattice monopole action $S_{mon}(k)$ and the
\emph{v.e.v.} of the squared magnetic charge, $\langle
k^2_s\rangle$. We study the quantity $\langle k^2_s\rangle$ instead of
the density, $\langle |k_s|\rangle$, since the analytical treatment of
the density is difficult while both these quantities are equivalent
philosophically.

\subsection{Lattice monopole action}

To derive the lattice monopole action we substitute the unity,
\begin{equation}
1 = \sum_{k(\Lambda) \in \Z} \prod_{s \in \Lambda} \delta_\Lambda
\left(k_s - \int\limits_{C_s} \dd^3 x \, \rho(x)\right),
\label{unity}
\end{equation}
into eq.~\eq{Z1}. Here $\sum_{k(\Lambda) \in \Z} \equiv \prod_{s \in
  \Lambda} \sum_{k_s \in \Z}$ and $\delta_\Lambda$ stands here for the
Kronecker symbol ($i.e.$, lattice $\delta$-function). We get
\begin{eqnarray}
\cZ  &=& \sum_{k(\Lambda) \in \Z} \sum\limits_{N=0}^\infty \frac{\zeta^N}{N!}
\Biggl[\prod\limits^N_{a=1} \int \dd^3 x^{(a)} \sum\limits_{q_a = \pm 1}\Biggr]
\int\limits^\pi_{-\pi} \dD_\Lambda h \int \dD \chi\times
\nonumber\\&&
\times\exp\left\{
- \int \dd^3 x \left[ \frac{1}{2 g^2_M} {\bigl( \partial_i \chi(x)\bigr)}^2
+ i \rho(x) \left(\chi(x) - \sum_{s \in \Lambda} \theta_s(x)
h_s\right)+ i \sum_{s \in \Lambda} k_s h_s\right]
\right\},\qquad
\label{Z2}
\end{eqnarray}
where we have introduced two additional integrations over the
continuum field $\chi$ and the compact lattice field $h$ to represent
the inverse laplacian in eq.~\eq{Z1} and the Kronecker symbol in
eq.~\eq{unity}, respectively. The subscript $\Lambda$ in $\dD_\Lambda
h$ indicates that the integration is going over the lattice fields
$h$. The representative \pagebreak[3] function of the $s^{\mathrm{th}}$ lattice cell
is denoted as $\theta_s$:
\begin{equation}
\theta_s(x) = \left\{
\begin{array}{ll}
1\,, & x \in C_s\,, \\
0\,, & \mbox{otherwise}\,. \\
\end{array}
\right.
\end{equation}

Summing over the continuum monopole positions according to
ref.~\cite{Polyakov}, we get
\begin{equation}
Z = \sum_{k(\Lambda) \in \Z} e^{- S_{mon}(k)}\,,
\end{equation}
where the lattice monopole action, $S_{\rm mon}$, is defined as follows:
\begin{eqnarray}
e^{- S_{\rm mon}(k)} & = & \int \dD \chi \int\limits^\pi_{-\pi} \dD_\Lambda h
\exp\Biggl\{ - \int \dd^3 x \Biggl[ \frac{1}{2 g^2_M} {\bigl( \partial_i \chi(x)\bigr)}^2-
\label{Z3}\\&&
         \hphantom{\int \dD \chi \int\limits^\pi_{-\pi} \dD_\Lambda h
\exp\Biggl\{- \int \dd^3 x \Biggl[}
- 2 \zeta \cos \left(\chi(x) - \sum_{s \in \Lambda} \theta_s(x)
h_s\right)+ i \sum_{s \in \Lambda} k_s h_s\Biggr] \Biggr\}\,.
\nonumber
\end{eqnarray}
The functional integral over the field $\chi$ in the l.h.s. of this
equation resembles the sin-Gordon model coupled to an external source.

Note that the left hand side of eq.~\eq{Z3} is invariant both under
the global continuum transformations of the field $\chi$, and under
the lattice local transformations of the field $h$,
\begin{eqnarray}
\begin{array}[b]{llll}
\chi(x) & \to & \chi(x) + 2 \pi n\,, &\quad n \in \Z \\
h_s & \to & h_s + 2 \pi m_s\,, & \quad m_s \in \Z\,,
\end{array}
\end{eqnarray}
respectively. Due to the last invariance we extend below the
integration over the lattice field $h$ to infinite limits (this leads
to appearance of an inessential factor in front of the partition
function).

Let us consider the lattice monopole action in the tree (or, gaussian)
approximation. The validity of this approximation is discussed in section~\ref{sec:validity}.
We expand the cosine function over the small fluctuations in $\chi$ and $h$ and we
get in a leading order:
\begin{eqnarray}
e^{- S^{\rm tree}_{\rm mon}(k)} & = & \int \dD \chi \int\limits^\pi_{-\pi} \dD_\Lambda h
\exp\Biggl\{ - \int \dd^3 x \Biggl[ \frac{1}{2 g^2_M} {\bigl( \partial_i \chi(x)\bigr)}^2+
\label{Z4}\\&&
         \hphantom{\int \dD \chi \int\limits^\pi_{-\pi} \dD_\Lambda h
\exp\Biggl\{ - \int \dd^3 x \Biggl[}
+ \zeta {\left(\chi(x) - \sum_{s \in \Lambda} \theta_s(x)
h_s\right)}^2 + i \sum_{s \in \Lambda} k_s h_s\Biggr] \Biggr\}\,.
\nonumber
\end{eqnarray}

The integration over field $\chi$ leads to the following expression:
\begin{equation}
e^{- S^{\rm tree}_{\rm mon}(k)} = \int\limits^\infty_{-\infty}
\dD_\Lambda h \exp\Biggl\{ - \zeta b^3 \sum_{s,s'} h_s \cF^{-1}_{s,s'}
h_{s'} + i \sum_{s \in \Lambda} k_s h_s \Biggr\}\,,
\label{Z5}
\end{equation}
where
\begin{eqnarray}
\cF^{-1}_{s,s'} & = & \delta_{s,s'} - m^2_D b^2 \cG_{s,s'}\,, 
\label{F}\\
\cG_{s,s'} & = & \frac{1}{b^5} \int \dd^3 x \int \dd^3 y \theta_s (x)
\, D_{m_D}(x-y) \, \theta_{s'}(y)\,, 
\label{G}
\end{eqnarray}
where $D_{m_D}$ is the scalar propagator for a massive particle, $( -
\partial^2_i + m^2) D_m(x) = \delta^{(3)} (x)$, with the Debye mass $m
= m_D \equiv \lambda^{-1}_D$. Note that the lattice operators $\cF$
and $\cG$ are dimensionless quantities.

In eq.~\eq{Z5} the integration over the lattice field $h$ can be
performed straightforwardly giving us the action for the lattice
monopoles:
\begin{equation}
S^{\rm tree}_{\rm mon}(k) = \frac{1}{4 \zeta b^3} \sum_{s,s'} k_s \,
\cF_{s,s'} k_{s'}\,.
\label{mon:tree}
\end{equation}

Let us calculate the operator $\cF$ on the infinite lattice.  We
represent the propagator $D_{m_D}$ as an integral over continuum
momentum and integrating over $x$ and $y$ coordinates in eq.~\eq{G}:
\begin{equation}
\cG_{s,s'} = \frac{1}{b^5} \int \frac{\dd^3 p}{(2 \pi)^3} \frac{1}{p^2
  +m^2_D} \, \Biggl[ \prod^3_{i=1} \Biggl( \frac{2 \sin (b p_i \slash
    2)}{p_i} \Biggr)\Biggr] e^{ i (s - s', p) b}\,,
\label{G2}
\end{equation}
where $(a,b)$ denotes the scalar product of the vector quantities.
Changing in eq.~\eq{G2} the integration variable, $p = q \slash b$,
and introducing the dimensionless mass,
\begin{equation}
\mu = b \slash \lambda_D\,,
\label{mu}
\end{equation}
we get
\begin{equation}
\cG_{s,s'} = \int \frac{\dd^3 q}{(2 \pi)^3} \, H(q,\mu) \, e^{ i (s -
  s', q)}\,, \qquad H(q,\mu) = \frac{1}{q^2 + \mu^2} \, \prod^3_{i=1}
   {\Biggl[ \frac{\sin (q_i \slash 2)}{q_i/2} \Biggr]}^2\,.
\label{H}
\end{equation}

Now we have to find the operator $\cF$ which is inverse to
eq.~\eq{F}. We represent $\cF$ as follows:
\begin{equation}
\cF_{s,s'} = \int^\pi_{-\pi} \dd^3 u \, \cF(u) \, e^{i (s - s', u)}\,,
\label{Fss}
\end{equation}
Substituting this equation and eq.~\eq{F} in the relation,
$$
\sum_{s'' \in \Lambda} \cF^{-1}_{s,s''} \cF_{s'',s'} = \delta_{s,s'}\,,
$$
we get the equation,
\begin{equation}
\int^\pi_{-\pi} \dd^3 u\, \left\{\cF(u) \left[ 1 - \mu^2 \sum_{r \in
    \Lambda} H(u + 2 \pi r,\mu)\right] - \frac{1}{(2 \pi)^3} \right\}
e^{i(s - s',u)} =0\,,
\label{eq:to:solve}
\end{equation}
where we have used the Poisson summation formula,
\begin{equation}
\sum_{s'' \in \Lambda} e^{i (s'', u - q)} = (2 \pi)^3 \, \sum_{r \in
  \Lambda}\delta^{(3)} (q - u - 2 \pi r)\,,
\end{equation}
and then integrated over the momentum $q$.

The solution to eq.~\eq{eq:to:solve} is
\begin{equation}
\cF(u) = \frac{1}{(2 \pi)^3} \frac{1}{1 - \mu^2 \sum_{r \in \Lambda}
  H(u+ 2 \pi r,\mu)}\,.
\label{Fu}
\end{equation}

Substituting eq.~\eq{Fu} in eq.~\eq{Fss} and using definition \eq{H},
we finally get
\begin{equation}
\cF_{s,s'} = \int\limits^\pi_{-\pi} \frac{\dd^3 u}{(2 \pi)^3} \,
{\Biggl[\sum_{r(\Lambda) \in \Z} \sum_{i=1}^3
\frac{4\, \sin^2 (u_i \slash 2)}{(\vec u + 2 \pi \vec r)^2 + \mu^2}\,
\prod^3_{\stackrel{j=1}{j \neq i}} {\Biggl(
\frac{2 \sin (u_j \slash 2)}{u_j + 2 \pi r_j} \Biggr)}^2
\Biggr]}^{-1} \!\!\!\!\!\cdot e^{i (s - s', u)}\,,
\label{Fss2}
\end{equation}
where we have used the relation
\begin{equation}
\sum_{r_i \in \Z} \frac{1}{{(u_i + 2 \pi r_i)}^2} = \frac{1}{[4\,
    \sin^2(u_i \slash 2)]}\,.
\label{rel:sum}
\end{equation}

The operator $\cF_{s,s'}$ coincides with a three-dimensional perfect
propagator for a free scalar field space-time discussed in details in
ref.~\cite{Wolfgang1}.

The summation in eq.~\eq{Fss2} over one of the integers $r_i$ can
explicitly be done with the help of the relation,
\begin{equation}
\sum_{r_i \in \Z} \frac{1}{(\vec u + 2 \pi \vec r)^2 + \mu^2} =
\frac{1}{2 \lambda_i(\vec u + 2 \pi \vec r,\mu)}
\frac{\sinh\lambda_i(\vec u + 2 \pi \vec r,\mu)}{ \cosh\lambda_i(\vec
  u + 2 \pi \vec r,\mu) - \cos u_i }\,,
\label{sum}
\end{equation}
where
\begin{equation}
\lambda_i(\vec q,\mu) = {\left[\sum^3_{\stackrel{j=1}{j \neq i}} q^2_j + \mu^2\right]}^{{1}/{2}}\,.
\label{lambda}
\end{equation}
Substituting eq.~\eq{sum} into eq.~\eq{Fss2} we get the final
expression for operator $\cF$:
\begin{eqnarray}
\cF_{s,s'} &=&\int\limits^\pi_{-\pi} \frac{\dd^3 u}{(2 \pi)^3} \, e^{i
  (s - s', u)}\times
\label{Fss3}\\&&
\times \left[\sum_{i=1}^3 \sum_{r_j \in \Z \atop j\neq i}
  \frac{2\, \sin^2 (u_i \slash 2)}{\lambda_i(\vec u + 2 \pi \vec
    r,\mu)} \frac{\sinh\lambda_i(\vec u + 2 \pi \vec r,\mu)}{
    \cosh\lambda_i(\vec u + 2 \pi \vec r,\mu) - \cos u_i} \cdot
  \prod^3_{k=1 \atop k \neq i} \frac{4 \sin^2 (u_k \slash
    2)}{(u_k + 2 \pi r_k)^2} \right]^{-1} \,.\nonumber
\end{eqnarray}
The leading term in the lattice monopole action is defined by
eqs.(\ref{mon:tree}), (\ref{Fss3}).

The finite-volume expression for the lattice monopole action can be
obtained from eq.~\eq{Fss3} by the standard substitution:
\begin{equation}
u_i \to \frac{2 \pi k_i}{L_i}\,, \qquad k =0\,,1\,,\dots L_i-1\,, \qquad
\int\limits^\pi_{-\pi} \frac{\dd u_i}{2\pi} \to \frac{1}{L_i}
\sum\limits_{k_i =0}^{L_i - 1}\,,
\label{finite:lattice}
\end{equation}
where $L_i$ is the lattice size in $i^{\mathrm{th}}$ direction.

In the infinite-volume case the lattice operator $\cF_{s,s'}$ depends
only on the dimensionless quantity $\mu$, eq.~\eq{mu}, which is the
ratio of the lattice monopole size $b$ and the Debye screening length,
eq.~\eq{lambdaD}. As we will see below the form of the operator $\cF$
is qualitatively different in the limits of small and large
$\mu$. Thus the Debye length $\lambda_D$ sets a scale for the lattice
monopole size (or, better to say, for the size of the lattice cell)
which characterizes different forms of the lattice monopole action.

\subsubsection{Action for large lattice monopoles}

Let us consider the case of the large lattice monopoles, $b \gg
\lambda_D$, or, equivalently, $\mu \gg 1$. In this case eq.~\eq{Fss3}
can be simplified since functions $\lambda_i$ (eq.~\eq{lambda}) are
large and $\sinh \lambda_i \sim \cosh \lambda_i \sim e^{\lambda_i}
\slash 2$.  Up to $O(e^{- \mu})$ corrections the operator $\cF_{s,s'}$
is given by
\begin{equation}
\cF_{s,s'} (\mu \gg 1) = \int\limits^\pi_{-\pi} \frac{\dd^3 u}{(2
  \pi)^3} \, {\left[\sum_{i=1}^3 \sum_{r_j \in \Z \atop j\neq
	i} \frac{2 \sin^2 (u_i \slash 2)}{\lambda_i(u,\mu)}
    \prod^3_{k=1 \atop k \neq i} \frac{4 \sin^2 (u_k \slash
      2)}{(u_k + 2 \pi r_k)^2} \right]}^{-1} \!\!\!\!\cdot e^{i (s -
  s', u)}\,.
\label{Fss4}
\end{equation}

In the case of large $\mu$, the function $\lambda_i$ in
eq.~\eq{lambda} is close to $\mu$ for all $r_i \ll \mu \slash (2
\pi)$. The deviation of $\lambda$ from $\mu$ becomes substantial only
for terms with large summation variable $r_i$, $r_i \gg \mu \slash (2
\pi)$. However such large $r_i$'s are in any case suppressed as $\sim
\prod_{j \neq i} {(2 \pi r_j)}^{-2}$ according to
eq.~\eq{Fss4}. Therefore the approximation $\lambda_i = \mu$ would
lead only to $O(\mu^{-2})$ corrections in eq.~\eq{Fss4}. Up to these
corrections eq.~\eq{Fss4} reads as follows:
\begin{equation}
\cF_{s,s'} (\mu \gg 1) = \frac{\mu}{2} \int\limits^\pi_{-\pi}
\frac{\dd^3 u}{(2 \pi)^3} \, {\left[\sum_{i=1}^3 \sin^2 \left(\frac{u_i}{2}\right)
\prod^3_{j=1 \atop j \neq i} \sum_{r_j \in \Z}
\frac{4 \sin^2 (u_j \slash 2)}{(u_j + 2 \pi r_j)^2}
\right]}^{-1} \!\!\!\!\cdot e^{i (s - s', u)}\,.
\label{Fss5}
\end{equation}

In turn, eq.~\eq{Fss5} can be simplified with the help of the
relation~\eq{rel:sum}, giving:
\begin{equation}
\cF_{s,s'} (\mu \gg 1) = 2 \mu \, D_{s,s'}\,,
\label{Flargeb}
\end{equation}
where $D_{s,s'}$ is the lattice laplacian:
\begin{equation}
D_{s,s'} = \int\limits^\pi_{-\pi} \frac{\dd^3 u}{(2 \pi)^3} \,
{\left[4 \sum\limits_{i=1}^3 \sin^2 \left(\frac{u_i}{2}\right)
    \right]}^{-1} \, e^{i (s - s', u)}\,.
\label{laplacian}
\end{equation}

Thus for large sizes of the lattice cells, $b \gg \lambda_D$, the
lattice monopole action \eq{mon:tree} is of the Coulomb-type:
\begin{equation}
S^{\rm tree}_{\rm mon}(k) = C(b)\,\sum\limits_{s,s'} k_s \, D_{s,s'}\,
k_{s'} + O\left(\left(\frac{\lambda_D}{b}\right)^4\right)\,, \quad b \gg
\lambda_D\,.
\label{Coul}
\end{equation}
The pre-Coulomb term $C$ can be expressed through the continuum
density of the mo\-no\-po\-les, $\rho$, the Debye screening length
$\lambda_D$ and the lattice spacing $b$ as
\begin{equation}
C(b) = \frac{1}{\lambda_D \rho} \cdot \frac{1}{b^2}\,.
\label{C}
\end{equation}
Here we used eq.~(\ref{mu}).

Thus in the limit of large lattice monopole sizes, $b$, the lattice
monopole action is of the Coulomb form.  The long-range nature of the
Coulomb interaction in the effective lattice action and the screening
effect in the underlying high temperature theory do not contradict
each other as one may think. The analog of this situation in continuum
can be represented by a similar Coulomb gas model~\eq{CoulombModel}
which contain the long-range interaction part in the
lagrangian. However, the interactions between monopoles lead to the
screening effect in various correlators in this model. Obviously, the
lattice Coulomb gas model \eq{Coul} must also possess the screening
effect despite the long range nature of the Coulomb term in the
action.

The coefficient in front of the Coulomb term, $C(b)$, scales as $C(b)
\sim b^{-2}$. Note that this form of scaling is a non-perturbative
effect. Indeed, naively one could expect that this coefficient has to
be proportional to the (squared) renormalized magnetic charge in three
dimensions, $C(b) \sim g^2_M$. The magnetic charge is inversely
proportional to the $3D$ electric charge, $g_M \sim g^{-1}_{E,3D}$. In
the leading (tree) order of dimensional reduction formalism the
electric charge is given by $g_{E,3D} = g_{4D}(\Lambda, T) \cdot
\sqrt{T}$, where $T$ is the temperature, $g_{4D}$ is the running $4D$
charge of the $\SU(2)$ gluodynamics and $\Lambda$ is the
renormalization scale. As a result, $C(b) \sim {[\,g^2_{4D}(\Lambda,T)
    \, T\,]}^{-1}$. The size of the lattice monopole, $b$, may enter
the above expression only in the form of the renormalization scale,
$\Lambda \sim 1 \slash b$. However, this would lead only to
logarithmic $b$-dependence of the coefficient $C(b)$. Thus the
$b^{-2}$ dependence of the pre-Coulomb coefficient is clearly of a
non-perturbative nature.

\subsubsection{Action for small lattice monopoles}

In the case of small lattice monopoles the leading term of the
operator $\cF$ for $b \ll \lambda_D$ follows immediately from
eq.~\eq{F}: $F_{s,s'} = \delta_{s,s'}$. Thus the monopole
action~\eq{mon:tree} becomes:
\begin{equation}
S^{\rm tree}_{\rm mon}(k) = M(b)\,\sum\limits_{s} k^2_s
+ O\left(\left(\frac{b}{\lambda_D}\right)^{2}\right)\,,
\qquad b \ll \lambda_D\,,
\label{Mass}
\end{equation}
where the coefficient
\begin{equation}
M(b) = \frac{1}{4 {\rho}} \cdot \frac{1}{b^3}\,,
\label{M}
\end{equation}
plays a role of the lattice monopole mass. Indeed, the leading term in
the action for $N$ small lattice monopoles with the charges $q= \pm 1$
is $S = M(b) \cdot N$.

In summary, we have established that the monopole action depends on
the ratio of the lattice cell and the continuum Debye screening mass,
$b \slash \lambda_D$:
\begin{eqnarray}
S_{\rm mon}(k) =
\left\{
\begin{array}{lll}
\displaystyle\frac{1}{4\rho} \cdot \frac{1}{b^3} \cdot  \sum\limits_{s} k^2_s + \cdots\,, &
\qquad b \ll \lambda_D\,; \\
\displaystyle\frac{1}{\rho\, \lambda_D} \cdot \frac{1}{b^2}
\cdot \sum\limits_{s,s'} k_s\, D_{s,s'}\, k_{s'} + \cdots\,,&
\qquad b \gg \lambda_D\,,
\end{array}
\right.
\label{TheorAction}
\end{eqnarray}
where $D_{s,s'}$ is the inverse laplacian on the lattice. Thus the
leading contribution to the lattice monopole action is given by the
mass (Coulomb) terms for small (large) lattice monopoles.

\subsection{Validity of the gaussian approximation}
\label{sec:validity}

As we will see below, from the point of view of numerical computations
the most interesting and reliable case corresponds to the large-$b$
monopoles. Here we briefly discuss the validity of the gaussian
approximation in this limit.  The detailed derivation will be given
elsewhere~\cite{in:preparation}.

We have truncated the cosine function in eq.~\eq{Z3} and performed the
gaussian integration over the continuum field~$\chi$ in eq.~\eq{Z4} to
get the effective action in the leading order for the lattice field
$h_s$. This action appears under the exponential function in
eq.~\eq{Z5}. The integration over the lattice fields~$h$ led us to the
Coulomb action~\eq{Coul} for the lattice monopoles in the large-$b$
limit. Let estimate the Coulomb action~\eq{Coul} coming from the next
order of the truncation of the cosine function in eq.~\eq{Z3}.

Expanding the cosine function in eq.~\eq{Z3} up to the fourth order
and treating the correction as a perturbation we arrive to the
following correction to the monopole action,
\begin{equation}
\delta S^{(4)}_{\rm mon}(k) = - \frac{\zeta}{12} \, \int \dd^3 x \,
\left\langle {\left(\chi(x) - \sum_{s \in \Lambda} \theta_s(x) h_s\right)}^4
\right\rangle,
\label{S4:1}
\end{equation}
where we used the superscript ``$(4)$" because, as we show below, this
action contains, $O(k^4)$ monopole-monopole interactions. The quantum
average~\eq{S4:1} is taken in the partition function~\eq{Z4}
containing the monopole current $k$ as a source.

Integrating the fields $\chi$ and $h$, and neglecting independent of
$k$ terms we arrive to the following $k^4$-correction to the monopole
action:
\begin{eqnarray}
\delta S^{(4)}_{\rm mon}(k) & = & - \frac{1}{192 \, \zeta^3 \, b^9}
\sum\limits_{s_1,\dots,s_4} k_{s_1} \dots k_{s_4} \, D_{s_1,\dots,
  s_4}\,,
\nonumber\\ 
D_{s_1,\dots, s_4} & =& \sum\limits_{s'_1,\dots,s'_4} \cF_{s_1,s'_1}
\dots \cF_{s_4,s'_4}\, Y_{s'_1,\dots, s'_4}\,,
\end{eqnarray}
where
\begin{eqnarray}
Y_{s'_1,\dots, s'_4} & = & {\left[ \prod\limits^4_{a=1} \int
    \frac{\dd^3 q^{(a)}}{{(2\pi)}^3} \frac{q^{(a),2}}{q^{(a),2} +
      \mu^2}\right]}\, {(2\pi)}^3 \, \delta^{(3)}\Bigl(q^{(1)}+ \cdots
+ q^{(4)}\Bigr)\times
\nonumber\\&&
\times{\left[ \prod^3_{i=1}\frac{\sin (q^{(a)}_i \slash
      2)}{q^{(a)}_i/2} \right]} \cdot \exp\left\{ i \sum^4_{b=1}
q^{(b)} s_b\right\}.
\label{Y}
\end{eqnarray}
The operator $Y$ is a converging function both in the limits of large
and small $q$.

According to eq.~\eq{Flargeb} the operator $\cF$ is proportional to
the first power of $b$ in the limit $\mu \equiv m_D b \gg 1$.  The
estimation of the operator $Y$, eq.~\eq{Y}, is difficult even in the
limit of large $b$.  However, one can notice that $Y$ contains the
dependence on $b$ in the denominators under the integrals over $q$,
and, similarly to the operator $\cF^{-1}$, the operator $Y$ should not
be a rising function of the scale $b$. Estimating $Y$ as being of the
order of unity, we get: $|S^{(4)}_{\rm mon}(k)| \lesssim g^4_M/(48
\zeta b^5) O(k^4)$. Thus, in the limit $b \to \infty$ the action
$S^{(4)}_{\rm mon} $ provides a small correction to the Coulomb
action~\eq{Coul}, which is proportional to $b^{-2}$. These results are
confirmed by the fact that in the limit of large monopoles $k^4$
correction to the monopole action is small compared to the quadratic
terms, ref.~\cite{NumericalMonopoleAction}. Moreover, as we will see
below, our numerical calculations shows that the quadratic part of the
monopole action can be described by the Coulomb interactions with a
high accuracy.

\subsection{Lattice monopole density}

In this section we discuss the dependence of the density of the
extended lattice monopoles on the monopole size $b$. The simplest
quantity characterizing the monopoles is the monopole density
$\rho_{latt}(b)$ measured in the \emph{lattice} units:
\begin{equation}
\rho_{\rm latt}(b) = \frac{1}{L^3} \, \left\langle \sum_{s \in \Lambda}
|k_s| \right\rangle,
\label{dens:latt}
\end{equation}
where $L$ is the lattice size in units of $b$.  However, the treatment
of eq.~\eq{dens:latt} from the analytical point of view is more
difficult than that of the squared density of the squared lattice
monopole charges:
\begin{equation}
\langle k^2 (b)\rangle =  \frac{1}{L^3} \, \left\langle
\sum_{s \in \Lambda} k^2(s) \right\rangle.
\label{dens2:latt}
\end{equation}
Below we discuss the quantity \eq{dens2:latt} which has a similar
physical meaning to the lattice monopole density~\eq{dens:latt}.

Both in eq.~\eq{dens:latt} and eq.~\eq{dens2:latt} the value of $k_s$
is equal to the total magnetic charge of the continuum monopoles which
are placed inside the corresponding cube of the volume
$b^3$. Obviously the lattice density \eq{dens:latt} does not give the
correct density of the continuum monopoles in general since if the
lattice monopole size is large enough then the oppositely charged
continuum monopoles cancel fields of each other inside the cell
$C_s$. Moreover, the dependence of the densities \eq{dens:latt} and
\eq{dens2:latt} on the lattice monopole size $b$ must reflect the
dynamics of the continuum monopoles. Indeed, one can expect that the
functions $\rho_{\rm latt}(b)$ for the Coulomb monopole gas and for
the random monopole ensembles should differ from each other since the
inter-monopole correlations are absent in the latter case contrary to
the former one. As we will see below the situation is similar to the
lattice monopole action discussed in the previous section.

Using eq.~\eq{ks} the lattice density \eq{dens2:latt} can be written
in the continuum theory as follows:
\begin{equation}
\langle k^2 \rangle = \int_{C_s} \dd^3 x \int_{C_s} \dd^3 y \, \langle
\rho(x) \, \rho(y) \rangle\,,
\label{d1}
\end{equation}
where the lattice site $s$ is fixed and the average is taken in the
Coulomb gas of the magnetic monopoles described by the partition
function \eq{Z1}. We assume the validity of the dilute gas
approximation for the continuum monopoles.

The correlator of the continuum monopole densities, $\langle \rho(x)
\, \rho(y) \rangle$, is well known from ref.~\cite{Polyakov}.
Introducing the source for the magnetic monopole density, $J$,
eq.~\eq{d1} can be rewritten as follows:
\begin{equation}
\langle k^2 \rangle = - \int_{C_s} \dd^3 x \int_{C_s} \dd^3 y \,
\frac{\delta^2}{\delta J(x) \, \delta J(y)} \left\langle\exp\left\{
i \int \dd^3 z \rho(z)\, J(z)\right\}\right\rangle{\biggl |}_{J=0}\,.
\end{equation}
Then we repeat the transformations in the previous section which led
us to eq.~\eq{Z4}. Integrating over quadratic fluctuations of the
field $\chi$ we get in the leading order
\begin{equation}
\langle \rho(x) \, \rho(y) \rangle = \rho \, \left[\delta^{(3)}(x-y) -
  m^2_D D_{m_D}(x-y)\right].
\label{corr}
\end{equation}

Substituting eq.~\eq{corr} in eq.~\eq{d1} and taking the integrals
over the cell $C_s$ we get
\begin{equation}
\langle k^2 \rangle = \rho\, b^3 \, P(\mu)\,,
\label{rho:b}
\end{equation}
where $\mu$ is the dimensionless Debye mass, eq.~\eq{mu} and in the
thermodynamical limit the function $P$ is
\begin{equation}
P(\mu) \equiv {(\cF^{-1})}_{0,0} (\mu) = 1 - \mu^2 \, \int
\frac{\dd^3\, q}{{(2\pi)}^3} \, H(q,\mu)\,.
\label{P}
\end{equation}
Here the inverse matrix $\cF_{s,s'}$ is given by eq.~\eq{F} and the
function $H$ is defined in eq.~\eq{H}.  The finite-volume analog of
eq.~\eq{P} can be easily written using the substitution
\eq{finite:lattice}.  \emph{Note that eq.~(\ref{rho:b}), (\ref{P})
  establish a direct relation between the density of the squared
  lattice monopole charges and the lattice monopole action,
  eq.~\eq{mon:tree}.}

It is interesting to study the scaling laws of the lattice monopole
densities for large and small monopoles. In the limiting cases the
behaviour of the function $P$ is as follows:
\begin{eqnarray}
P(\mu) = \left\{
\begin{array}{ll}
C_1 \, \mu^{-1} +
O(\mu^{-3})\,, & \qquad\mu \to \infty\,, \\
1 - C_2 \mu^2 + O(\mu^4)\,, & \qquad \mu \to 0\,,
\end{array}
\right.
\label{limits}
\end{eqnarray}
where the constants $C_{1,2}$ are
\begin{equation}
C_1 \approx 2.94\,, \qquad C_2 = \int \frac{\dd^3 q}{{(2\pi)}^3} \,
H(q,0) \approx 0.148\,.
\end{equation}

Substituting the asymptotic functions \eq{limits} in eq.~\eq{rho:b} we
get the scaling laws (in physical units) for the density of the
squared lattice monopole charge:
\begin{eqnarray}
\langle k^2 \rangle = \left\{
\begin{array}{ll}
C_1 \, \rho \, \lambda_D \, b^2 \cdot \left[1 +
O\left({\left(\lambda_D \slash b \right)}^2\right)\right]\,, & \qquad b \gg \lambda_D\,, 
\\
\rho \, b^3 \cdot \left[1 + C_2 \, \rho \, {(b \slash \lambda_D)}^2\,
+ O\left({\left( b \slash \lambda_D\right)}^4\right)\right]\,, & \qquad b \ll
\lambda_D\,.
\end{array}
\right.
\label{dens:limits}
\label{TheorDensity}
\end{eqnarray}

One can note some interesting properties of the density of the squared
magnetic charge, eq.~\eq{dens:limits}.
\begin{enumerate}
\item[(i)] The dependence of the lattice monopole charge squared on
  the monopole size $b$ is always polynomial, $b^n$. In particular,
  this property is remarkable in the large-$b$ region: one may expect
  that the screening of the monopole charges in the magnetic monopole
  plasma would lead to an exponential behaviour, $e^{- \mbox{const.}\,
    m_D \, b}$, which is not the case.
\item[(ii)] As in the case of the lattice monopole action, the power
  $n$ of the leading scaling law depends crucially on the value of the
  ratio of the lattice monopole size and the Debye screening length.
\item[(iii)] The proportionality of the density $\langle k^2 \rangle$
  to $b^2$ in large-$b$ region has a simple explanation. In a random
  gas of continuum monopoles we would get $\langle k^2 \rangle \sim
  \rho \, b^3$ since the monopoles are not correlated with each
  other. In the Coulomb gas the monopoles are correlated and moreover,
  screened. Therefore the monopoles separated from the boundary of the
  cell by the distance larger than $\lambda_D$, do not contribute to
  $\langle k^2 \rangle$. Consequently, the $b^3$ proportionality for
  the random gas turns into $\lambda_D b^2$ in the Coulomb gas and we
  get $\langle k^2 \rangle \sim \rho \, \lambda_D \, b^2$. The
  coefficient of proportionality is of a geometrical origin.
\item[(iv)] In the small $b$ region the density of the squared lattice
  monopole charges is equal to the density of the continuum monopoles
  times the volume of the cell. This is natural, since the smaller
  volume of the lattice cell, $b^3$, the smaller chance for two
  continuum monopoles to be located at the same cell. Therefore each
  cell predominantly contains not more that one continuum monopole,
  which leads to the relation $k^2_s = |k_s| = 0,1$. As a result we
  get $\langle k^2 \rangle \to \rho_{\rm latt}(b) \to \rho \, b^3$ in
  the limit $b \to 0$.
\end{enumerate}

Closing this section we mention interesting relations between the
density of the small- and large- sized lattice monopoles and the
coefficients in front of, respectively, the mass and the Coulomb terms
of the monopole action, eqs.(\ref{TheorAction}), (\ref{TheorDensity}):
\begin{equation}
C(b) = \frac{C_1}{\langle k^2 (b)\rangle}\,, \qquad b \gg \lambda_D\,,
\qquad \hbox{and} \qquad M(b) = \frac{1}{4 \, \langle k^2
  (b)\rangle}\,, \qquad b \ll \lambda_D\,.
\label{CM}
\end{equation}
All these results are obtained in the gaussian approximation.  The
results on next to the leading order correction to the action and the
monopole density will be published elsewhere~\cite{in:preparation}.

\section{Numerical results}
\label{experiment}

\subsection{Details of simulations}

In order to get perfect lattice monopole action and density we perform
numerically blockspin transformations for the lattice monopole
currents. The original model is define on the fine lattice with the
lattice spacing $a$ and after the blockspin transformation, the
renormalized lattice spacing becomes $b=na$, where $n$ is the number
of steps of blockspin transformations. The continuum limit is taken as
the limit $a \to 0$ and $n \to \infty$ for a fixed physical scale $b$.

Finite temperature system possesses a periodic boundary condition for
time direction and the physical length in the time direction is
limited to less than $1/T$.  In this case it is useful to introduce
anisotropic lattices.  In the space direction, we perform the
blockspin transformation and the continuum limit is taken as $a_s \to
0$ and $n_s \to \infty$ for a fixed physical scale $b=n_s a_s$.  Here
$a_s$ is the lattice spacing in the space directions and $n_s$ is the
blockspin factor.  In the time direction, the continuum limit is taken
as $a_t \to 0$ and $N_t \to \infty$ for a fixed temperature $T=1/(N_t
a_t)$.  Here $a_t$ is the lattice spacing in the time direction and
$N_t$ is the number of lattice sites for the time direction.  After
taking the continuum limit, we finally get the effective lattice
monopole action which depends on the physical scale $b$ and the
temperature $T$.

In this paper the numerical procedure to generate the field
configurations is identical to the one used
ref.~\cite{NumericalMonopoleAction}. While referring an interested
reader to the above paper, here we mention for completeness briefly
the basic points of our numerical procedure.  The anisotropic Wilson
action for pure four-dimensional $\SU(2)$ QCD is written as
\begin{eqnarray}
S &=& \beta \left\{ \frac{1}{\gamma} \sum_{s, i > j \ne 4} P_{ij} (s)
+ \gamma \sum_{s, i \ne 4} P_{i4} (s) \right\} , 
\nonumber\\ 
P_{\mu \nu} (s) &\equiv& \frac{1}{4} \Tr \left[ \bbbone - U_\mu (s)
  U_\nu (s + \hat{\mu}) U_{\mu}^{\dagger} (s + \hat{\nu})
  U_{\nu}^{\dagger} (s) \right] + \hbox{h.c.}
\end{eqnarray}
If $\gamma=1$, the lattice is isotropic $(a_s = a_t)$.  We can
consider various lattice spacings $a_s$ and $a_t$ by varying the
parameters $\beta$ and $\gamma$ in the action.  The procedure to
determine the relation between the lattice spacings $a_s$, $a_t$ and
the parameters $\beta$, $\gamma$ is described in details in
ref.~\cite{NumericalMonopoleAction}. Using the parameters which are
determined in ref.~\cite{NumericalMonopoleAction}, we simulate the
pure four-dimensional $\SU(2)$ QCD on the lattice $48^3 \times L_t$
with $L_t = 12, 10, 8 ,6, 4$ corresponding to temperature $T/T_c =
1.6, 1.92, 2.4, 3.2, 4.8$.  The parameters used here are summarized in
Table~\ref{bgaa}.  The scaling behaviors for time-like lattice
monopole action at above mentioned lattices show in large $b$
region~\cite{NumericalMonopoleAction}.

\TABLE[t]{\begin{tabular}{|c|c|c|c||c|c|c|c|} \hline
  $\beta$  &  $\gamma$  &   $a_s$  &   $a_t$  &  $\beta$  &  $\gamma$  &   $a_s$  &   $a_t$  \\ \hline
   2.470   &   2.841    &   0.250  &   0.075  &   2.565   &   2.152    &   0.180  &   0.075  \\ \hline
   2.500   &   2.615    &   0.225  &   0.075  &   2.573   &   2.098    &   0.175  &   0.075  \\ \hline
   2.533   &   2.354    &   0.200  &   0.075  &   2.581   &   2.042    &   0.170  &   0.075  \\ \hline
   2.548   &   2.256    &   0.190  &   0.075  &   2.598   &   1.927    &   0.160  &   0.075  \\ \hline
\end{tabular}%
\caption{Parameters $\beta, \gamma$ and corresponding lattice
spacings $a_s, a_t$ in units of the zero-temperature string
tension, $\sqrt{\sigma_{T=0}}$.\label{bgaa}}}

To study the abelian monopole dynamics we generate the thermalized
non-abelian link fields $\{ U_\mu (s)\}$ and we perform abelian
projection in the Maximally abelian (MA) gauge~\cite{MaA} for each
$\SU(2)$ configuration. The MA gauge fixing condition is the
maximization of the quantity $R$,
\begin{equation}
\max\limits_{\Omega} R[U^{(\Omega)}]\,,\qquad R[U] = {\Tr} \sum_{s,
  \mu} \left[ U_\mu(s) \sigma_3 U_{\mu}^{\dagger}(s+\hat{\mu})
  \sigma_3 \right] .
\label{MaA:condition}
\end{equation}
under the $\SU(2)$ gauge transformations, $U \to U^{(\Omega)} =
\Omega^\dagger U \Omega$. The gauge fixing condition
\eq{MaA:condition} is invariant under an abelian subgroup of the group
of the $\SU(2)$ gauge transformations. Thus the condition
\eq{MaA:condition} corresponds to the partial gauge fixing, $\SU(2)
\to \UU(1)$.

After the MA gauge fixing, the abelian, $\{ u_\mu (s)\}$, and
non-abelian $\{ \tilde{U}_\mu (s)\}$ link fields are separated,
$\tilde{U}_\mu (s) = C_\mu (s) u_\mu (s)$ , where
\begin{equation}
C_\mu (s) = \pmatrix{ \sqrt{1-|c_\mu (s)|^2} & -c_{\mu}^{\ast}(s) \cr
  c_{\mu}(s) & \sqrt{1-|c_\mu (s)|^2}},\qquad u_\mu (s) =\pmatrix{
  e^{i \theta_\mu (s)} & 0 \cr 0 & e^{-i \theta_\mu (s)}}.
\end{equation}
The vector fields $C_\mu (s)$ and $u_\mu (s)$ transform like a charged
matter and, respectively, a gauge field under the residual $\UU(1)$
symmetry. Next we define a lattice monopole current (DeGrand-Toussaint
monopole)~\cite{DGT}. Abelian plaquette variables $\theta_{\mu\nu}(s)$
are written as
\begin{equation}
\theta_{\mu\nu}(s) = \theta_\mu (s) + \theta_\nu (s+\hat{\mu}) -
\theta_\mu (s+\hat{\nu}) - \theta_\nu (s) \,, \qquad ( -4\pi <
\theta_{\mu\nu}(s) \le 4\pi ) \,.
\end{equation}
It is decomposed into two terms using integer variables
$n_{\mu\nu}(s)$:
\begin{equation}
\theta_{\mu\nu}(s) \equiv \bar{\theta}_{\mu\nu}(s) + 2\pi
n_{\mu\nu}(s)\,, \qquad ( -\pi < \bar{\theta}_{\mu\nu}(s) \le \pi ) .
\end{equation}
Here $\bar{\theta}_{\mu\nu}(s)$ is interpreted as an electromagnetic
flux through the plaquette and $n_{\mu\nu}(s)$ corresponds to the
number of Dirac string piercing the plaquette.  The lattice monopole
current is defined as
\begin{equation}
k_\mu (s) = \frac{1}{2} \epsilon_{\mu\nu\rho\sigma} \partial_\nu
n_{\rho\sigma}(s+\hat{\mu})\,.
\label{k}
\end{equation}
It satisfies the conservation law $\partial_{\mu}^{\prime} k_\mu (s) =
0 $.

\pagebreak[3]

To study numerically the lattice monopoles of various lattice sizes we
use the so-called extended monopole
construction~\cite{ExtendedMonopoles}. At zero temperature the
extended monopoles can be defined in the symmetric way. They have the
physical size $b^3$ where $b = n a$.  The charge of the $n$-blocked
monopole is equal to the sum of the charges of the elementary lattice
monopoles inside the $n^3$ lattice cell. At finite temperature the
blocking of spatial and temporal currents should be done
separately~\cite{NumericalMonopoleAction}:
\begin{eqnarray}
K_{\mu \ne 4} (s_s , s_4) &=& \sum_{i , j = 0}^{n_s - 1} \sum_{l =
  0}^{n_t -1} k_{\mu \ne 4} \left( n_s s_s + ( n_s -1 ) \hat{\mu} + i
\hat{\nu} + j \hat{\rho}, n_t s_4 + l \right) ,
\nonumber\\
K_{4} (s_s , s_4) &=& \sum_{i , j , l = 0}^{n_s - 1} k_{4} \left( n_s
s_s + i \hat{\mu} + j \hat{\nu} + l \hat{\rho} , n_t s_4 + ( n_t -1)
\right) ,
\end{eqnarray}
where $n_s$ ($n_t$) is the number of blocking steps in space (time)
direction.

We consider only the $n_t=1$ case since we are interested in high
temperatures for which the monopoles are almost static. The lattice
blocking is performed only in spatial directions, $n_s=1 \dots 8$, and
we study only the static components $K_4$ among the $4D$ monopole
currents $K_\mu$ (below we denote $K_4$ as $k$.).  At high temperature
we disregard the spatial currents $K_i$ since they are not interesting
from the point of view of the long-range non-perturbative spatial
physics. The size of the lattice monopoles is measured (unless
otherwise specified) in terms of the zero temperature string tension,
$\sigma_{T=0}$.

In order to figure out whether the continuum monopole currents can be
described by the Coulomb gas model we should compare Monte Carlo
results with the appropriate analytical predictions derived in the
previous section. In principle this should allow us to obtain all
independent parameters of the Coulomb gas (the continuum monopole
density and the Debye screening length). However, the lattice
monopoles of small sizes are largely affected by the lattice artifacts
since in this case the number of blocking steps is small and possible
magnetic charges of such monopoles are restricted due to peculiarities
of the DeGrand-Toussaint definition~\cite{DGT}.  Moreover, the
short-range interaction between the continuum monopoles should deviate
from the simple Coulomb law due to a non-zero finite radius of the
abelian monopole~\cite{FiniteRadius}. Thus in order to get reliable
results we perform the comparison of the numerical data with
analytical predictions for sufficiently large blocking steps only. The
restriction to large-$b$ region allows us to calculate the product of
the continuum monopole density and the Debye screening length,
\begin{equation}
R(T) = \lambda_D(T)\,\rho(T)\,,
\label{R}
\end{equation}
\looseness=1while a separate calculation of these quantities is not possible.
Nevertheless the knowledge of quantity is enough to make a conclusion
about realization of the Coulomb gas picture for static continuum
monopole lines.  In subsections~\ref{sec:NumMonopoleAction} and
\ref{sec:NumMonopoleDensity} we get the quantity $R$ from the lattice
monopole action and density, respectively. To avoid confusion we
denote $R$ in these cases as $R_{act}$ and $R_\rho$,
correspondingly. In subsection \ref{sec:NumCheck} we check the
consistency of the obtained values of $R_{act}$ and $R_\rho$ with each
other and with the Coulomb gas picture.

\pagebreak[3]

\subsection{Lattice monopole action}
\label{sec:NumMonopoleAction}

\FIGURE{\epsfig{file=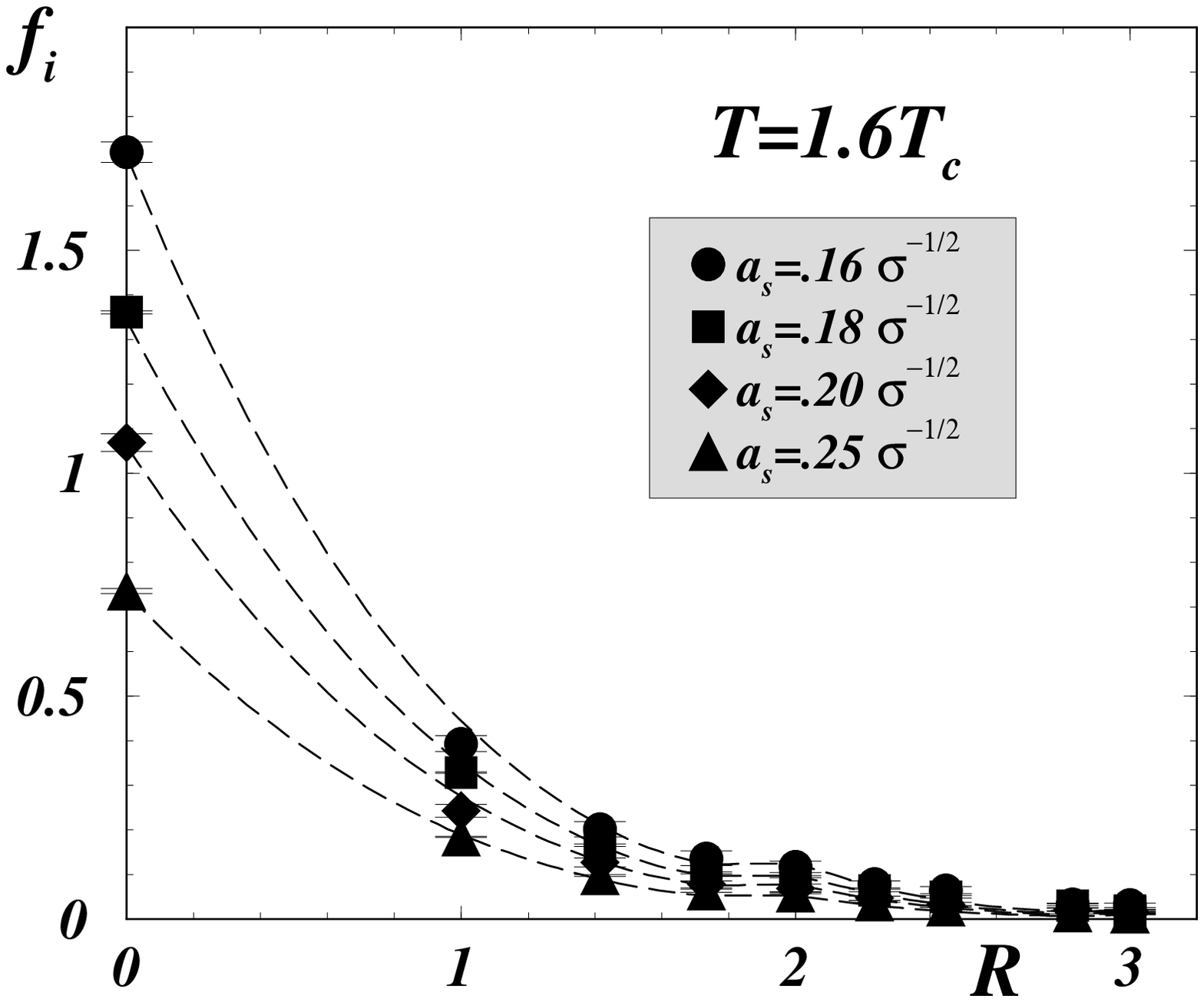,width=6cm}%
\caption{The coupling constants, $f_i$, of the lattice monopole action
  $vs.$ the distance, $r$ at $T=1.6 T_c$ for various spacings of the
  fine lattice, $a_s$.  The fits~\eq{CoulombLaw} are visualized by the
  dashed lines.\label{fig:fit:act1}}}

The lattice monopole action for the static monopole currents, $k_{s}$,
at high tem\-pe\-ra\-tu\-res was found numerically in
ref.~\cite{NumericalMonopoleAction} using an inverse Monte-Carlo
procedure. It turns out that the self-in\-ter\-ac\-tion of the
temporal lattice currents can be successfully described by the
quadratic monopole action:
\begin{equation}
S_{\rm mon}(k) = \sum_i f_i \, S_i(k)\,,
\end{equation}
where $S_i$ are two-point operators of the lattice monopole charges
corresponding to different separations between the charges. The term
$S_1$ corresponds to the zero distance between the lattice monopoles,
$S_2$ corresponds to the unit distance and so on (see
ref.~\cite{NumericalMonopoleAction} for further details).

The two-point coupling constants, $f_i$, of the lattice monopole
action are shown in figures~\ref{fig:fit:act1} and~\ref{fig:fit:act2}
as a function the distance between the lattice points. The numerical
data corresponds to lowest, $T=1.6 T_c$, and highest, $T=4.8 T_c$,
available temperatures. The number of blocking steps is fixed to
$n_s=6$ while the spatial spacings of the fine lattice, $a_s$, are
available for a set of values ranging from $a_s=.16 \sigma^{- 1 \slash
  2}$ till $a_s = .25 \sigma^{- 1 \slash 2}$.

\FIGURE{\epsfig{file=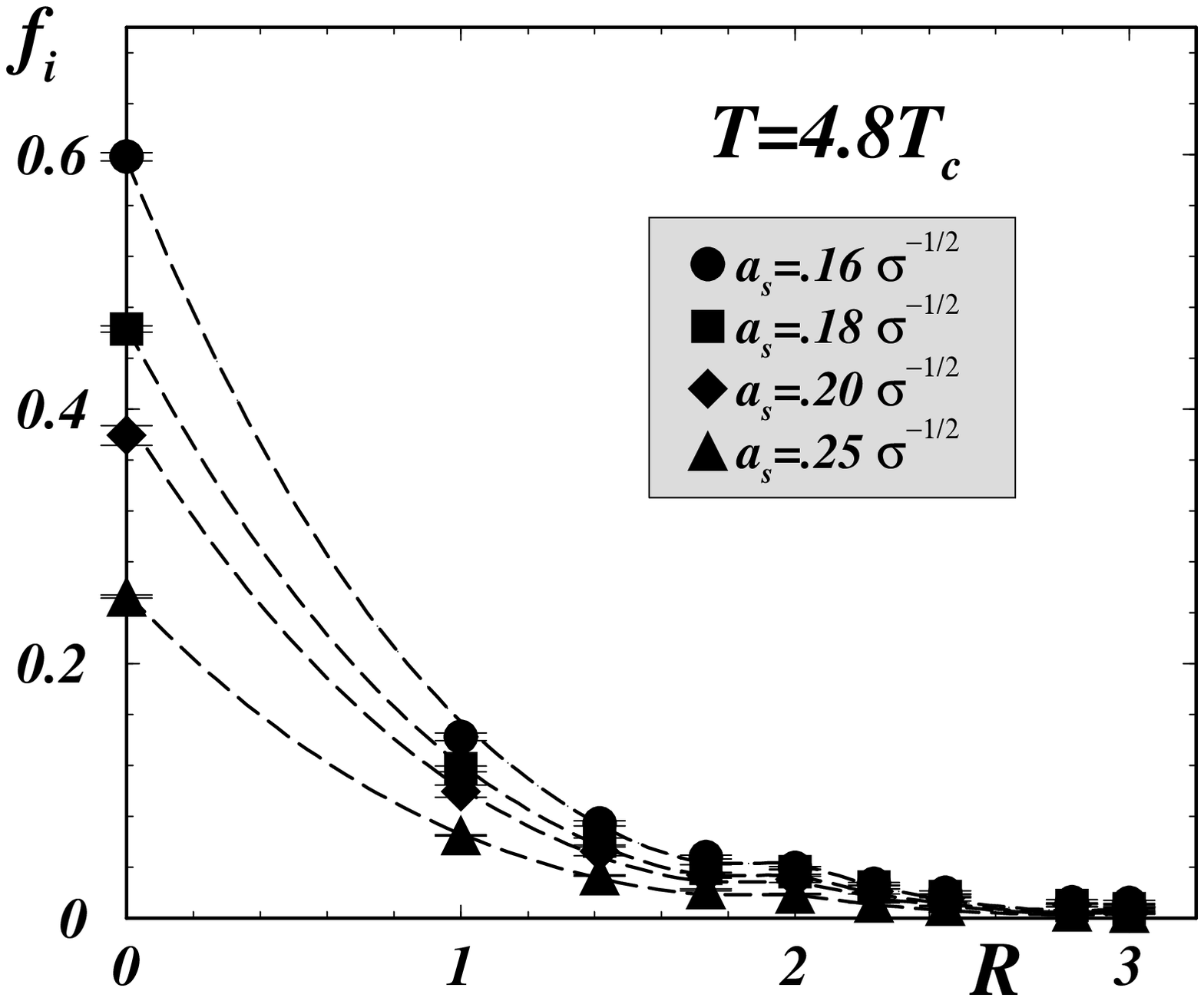,width=6cm}%
\caption{The same as in figure~\ref{fig:fit:act1} but for the
  temperature $T=4.8 T_c$.\label{fig:fit:act2}}}

According to eq.~\eq{TheorAction} the leading term in the lattice
monopole action for large lattice monopoles  ($b \gg
\lambda_D$) must be proportional to the Coulomb interaction,
\begin{equation}
S_{\rm mon}(k) = C_C \cdot \sum\limits_{s,s'} k_s\, D_{s,s'}\,
k_{s'}\,.
\label{CoulombLaw}
\end{equation}
To check this prediction we fit the coupling con\-stan\-ts $f_i$ by
the Coulomb interaction~\eq{CoulombLaw} treating $C_C$ is the fitting
parameter. The fits are visualized by the dashed lines in
figures~\ref{fig:fit:act1} and~\ref{fig:fit:act2}.

As one can see from the figures, this \emph{one-pa\-ra\-met\-ric} fit
works very good.  The $\chi^2/d.o.f.$ parameter is of the order of
unity for most fits while it is of order of 2 for the smallest lattice
spacing, $a \sqrt{\sigma} = 0.16$. Note that the dependence of the
lattice Coulomb interaction on the distance between the interaction
points is not a monotonic function, as one can see from the fitting
curves. Similar non-monotonic behaviour is also observed in the
numerical data.

\pagebreak[4]

Using the fitting of the action we obtain the values of $C_C$ for
various sizes of the lattice monopole, $b \sqrt{\sigma} = .96 \dots
1.5$ at different temperatures, $T = (1.6 \dots 4.8) T_c$.  According
to eq.~\eq{TheorAction} the pre-Coulomb coefficient $C_C(b,T)$ at
sufficiently large lattice monopole sizes, $b$ ($b \gg \lambda_D$)
should be as follows:
\begin{equation}
C_C(b,T) = \frac{1}{R(T) \, b^2}\,,
\label{PreCoulombAction}
\end{equation}
where $R$ is the product of the screening length and the continuum
monopole density, eq.~\eq{R}.

We present the data for the pre-Coulomb coefficient, $C_C(b,T)$ and
the corresponding \emph{one-pa\-ra\-me\-ter}
fits~\eq{PreCoulombAction} in figure~\ref{fig:cc1}. The fit is
one-parametric with $R$ being the fitting parameter. Again we observe
that the agreement between the data for $C_C$ and the fits is very
good.  We show the quantity $R$ $vs.$ temperature in
figure~\ref{fig:cc2}.

\DOUBLEFIGURE[t]{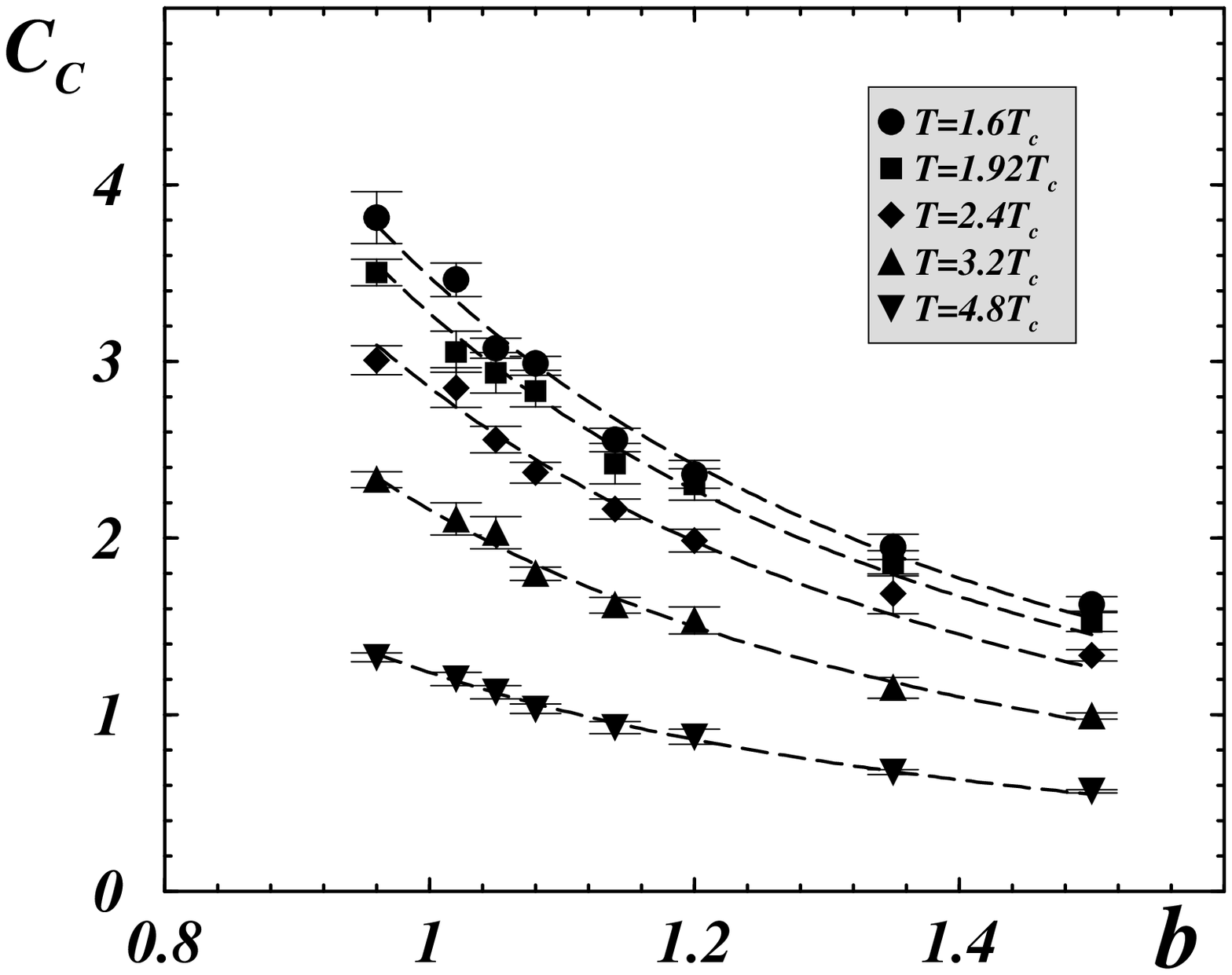,width=6.0cm}%
{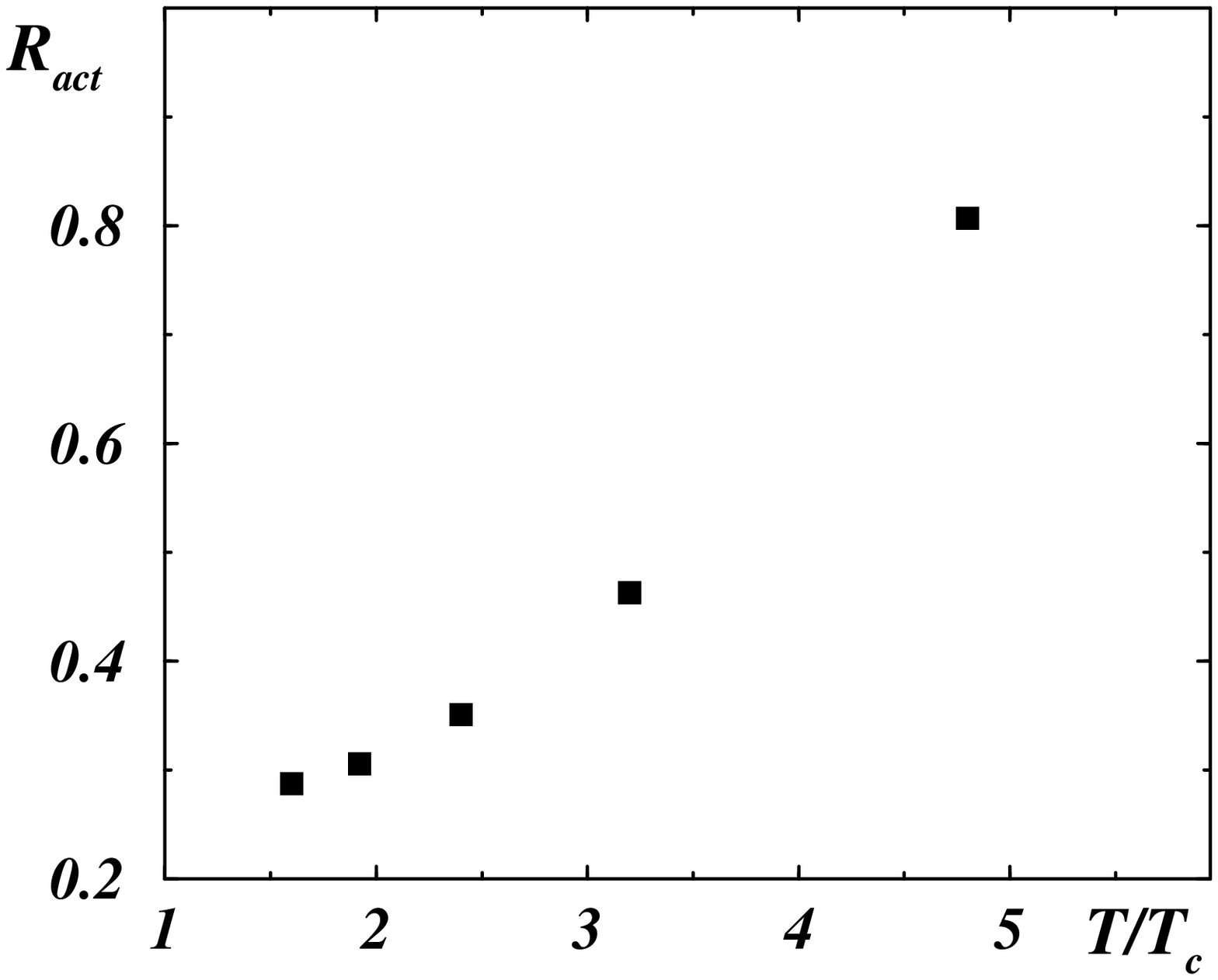,width=5.9cm}
{The pre-Coulomb coupling $C_C$ and the fits of $C_C$ by
  eq.~\protect\eq{PreCoulombAction} for various
  temperatures~$T$.\label{fig:cc1}}
{The product of the screening length and the continuum monopole
  density, eq.~\protect\eq{R}, calculated from the lattice monopole
  action (in units of the string tension).\label{fig:cc2}}

The lattice Coulomb form of the action and proportionality of the
pre-Coulomb term to $b^{-2}$ at large values of $b$ was
established~\cite{GG} also in the three-dimensional Georgi-Glashow
model for the 't~Hooft-Polyakov monopoles. These facts does not come
unexpected from the point of view of the discussion above.

\subsection{Lattice monopole density}
\label{sec:NumMonopoleDensity}

\looseness=1As we have seen from previous section the density of squared lattice
monopole charge should also contain the information about the
parameters of the Coulomb gas model. According to
eq.~\eq{TheorDensity} the large-$b$ asymptotics of the quantity
$\langle k^2(b)\rangle \slash b^2$ can be used to extract the product
of the screening length and the continuum monopole density $R$,
eq.~\eq{R}. We have measured the density of squared lattice monopole
charge for all available temperatures and lattice monopole sizes. As
an example we plot in figures~\ref{fig:p2:mon:dens1}
and~\ref{fig:p2:mon:dens2} the quantity $\langle k^2(b)\rangle \slash
b^2$ $vs.$ the lattice monopole size, $b$, for lowest and highest
available temperatures, respectively.

\DOUBLEFIGURE[t]{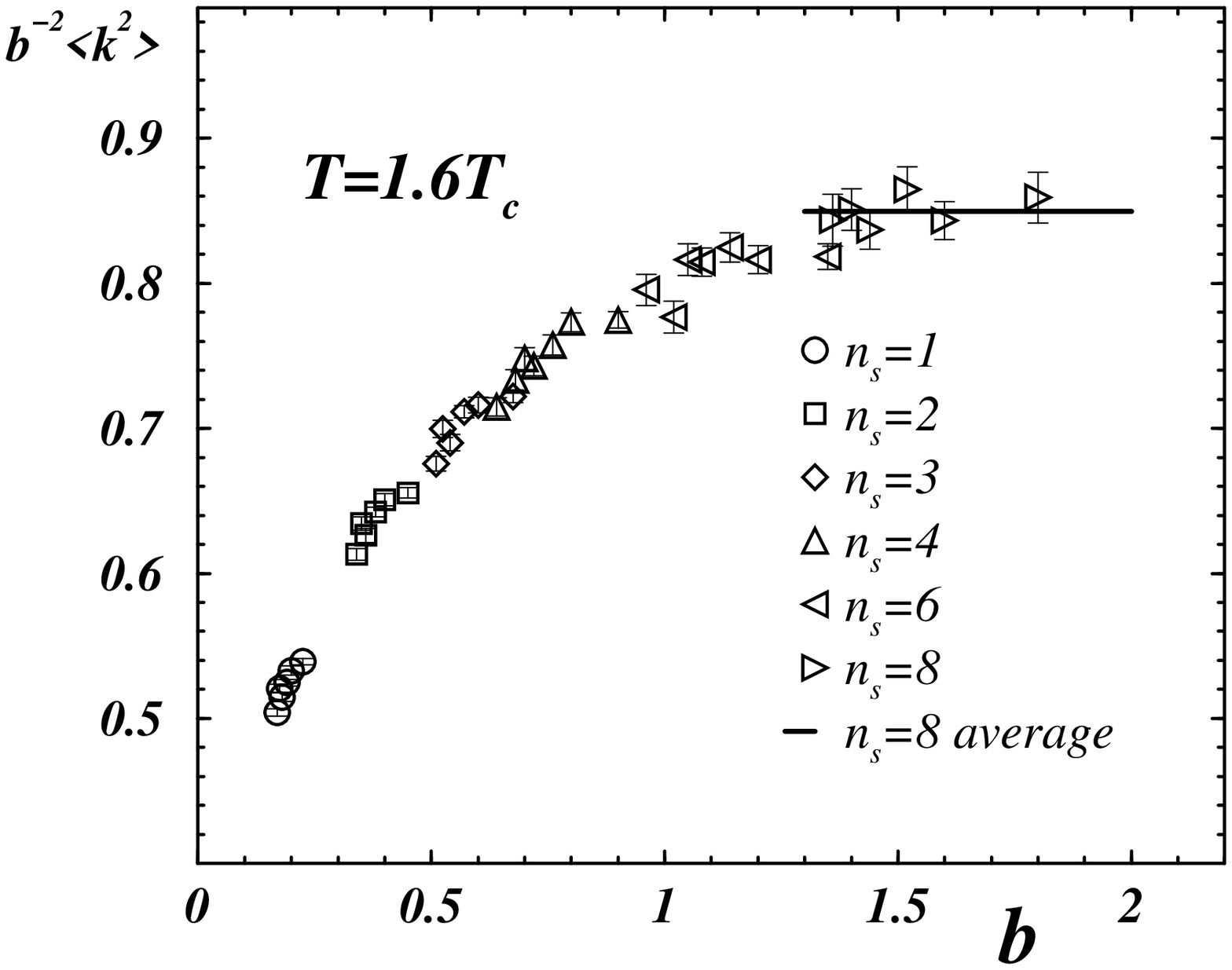,width=5.9cm}
{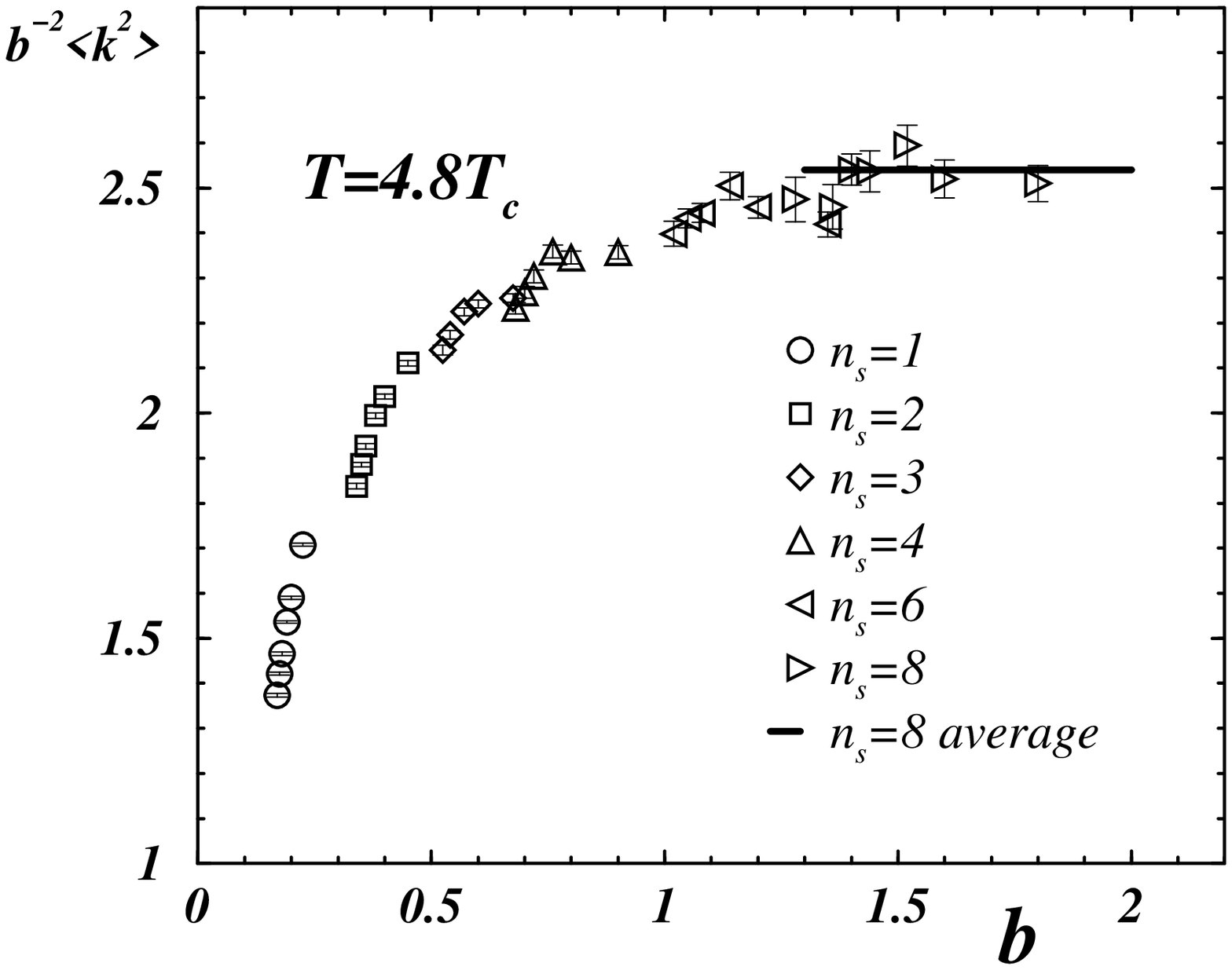,width=6.0cm}
{The ratio $\langle k^2 (b)\rangle \slash b^2$ vs.\ lattice monopole
  size, $b$, at $T = 1.6 \, T_c$. \label{fig:p2:mon:dens1}}
{The same as in figure~\protect\ref{fig:p2:mon:dens1}
but for $T = 4.8 \, T_c$.\label{fig:p2:mon:dens2}}

Our theoretical expectations~\eq{TheorDensity} indicate that the
function $\langle k^2 (b)\rangle \slash b^2$ must vanish at small
monopole sizes and tend to constant at large $b$. This behaviour can
be observed\pagebreak[3] in our numerical data, figure~\ref{fig:p2:mon:dens2}, up
to some jumps for densities with different $n_s$. We ascribe these
jumps to the lattice artifacts emerged due to finiteness of the fine
lattice spacing, $a$, and finite volume effects.

\FIGURE{{\epsfig{file=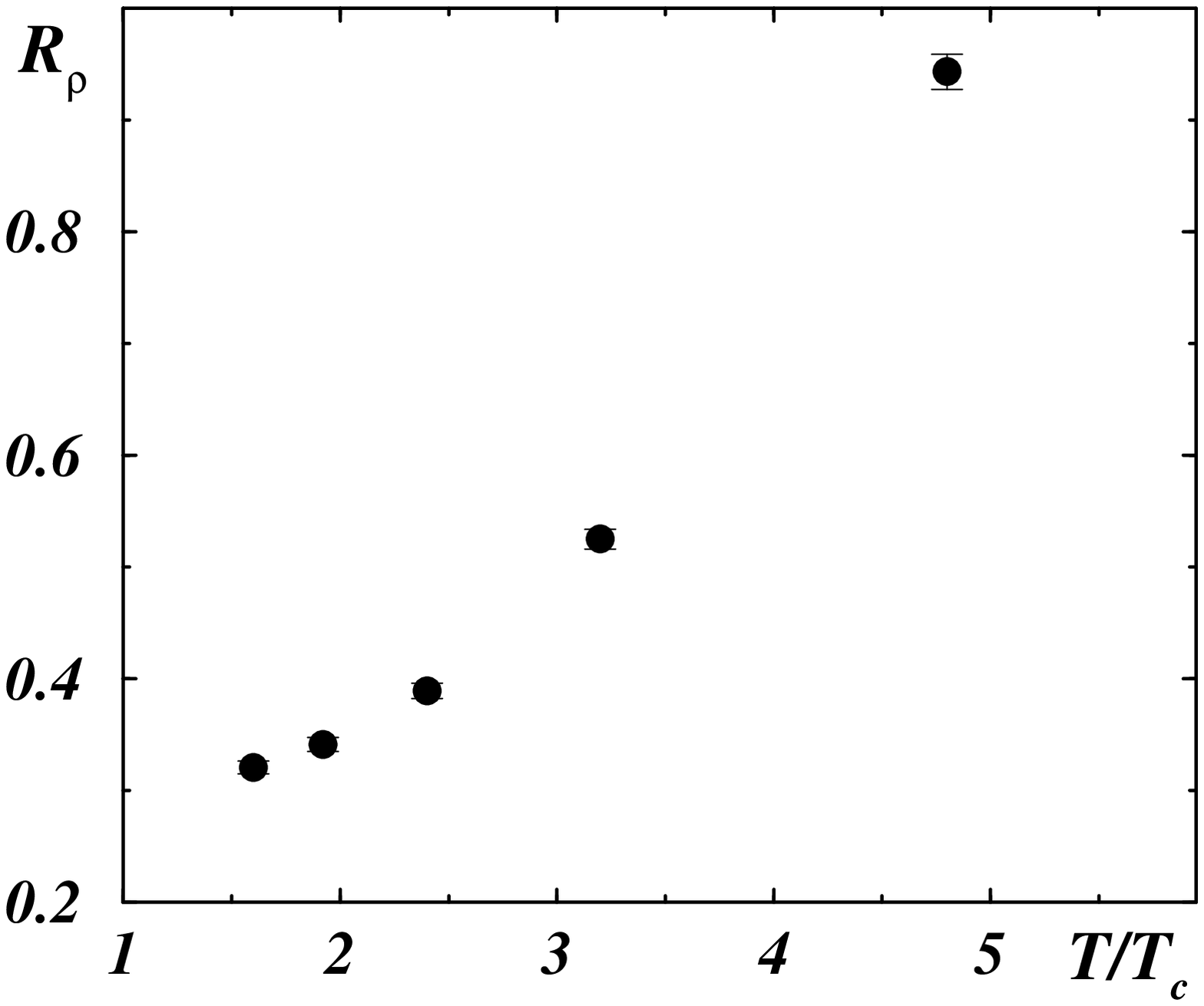,width=6cm}}
\caption{The same quantity
  as in figure~\ref{fig:cc2} but now calculated from the lattice
  monopole density.\label{fig:RhoLambdaDensity}}}

According to eq.~\eq{TheorDensity} we should know the large-$b$
asymptotics of $\langle k^2 (b)\rangle \slash b^2$ to get the quantity
$R$, eq.~\eq{R}. These asymptotics are approximated by averaging of
the appropriate $n_s=8$ data for which the behaviour of the function
in question is almost flat. Then we get the product of the screening
length and the continuum monopole density, $R$, depicted in
figure~\ref{fig:RhoLambdaDensity}.

The result is very similar to the one obtained from the behavior of
the lattice monopole action. However, the quantity $R$ obtained from
the lattice monopole density is a bit larger than the same quantity
calculated from the lattice monopole action. This fact can be expected
since we have approximated the asymptotics of $\langle k^2 (b)\rangle
\slash b^2$ by the average of the $n_s=8$ data which may slightly
differ from a correct asymptotics.

\subsection{Check of the Coulomb gas picture for continuum monopoles}
\label{sec:NumCheck}

Although both quantities $R_\rho$ and $R_{act}$ correspond to the
product of the screening length and the continuum monopole density,
from a numerical point of view both $R$'s are independent. To check
the self-consistency of our approach we plot the ratio of these
quantities in figure~\ref{fig:self1}. It is clearly seen that the
ratio is independent of the temperature and very close to unity, as
expected.  The 10\%--15\% deviation of this quantity from unity may be
explained by reasons mentioned in the end of the previous subsection.

\pagebreak[3]

\DOUBLEFIGURE[t]{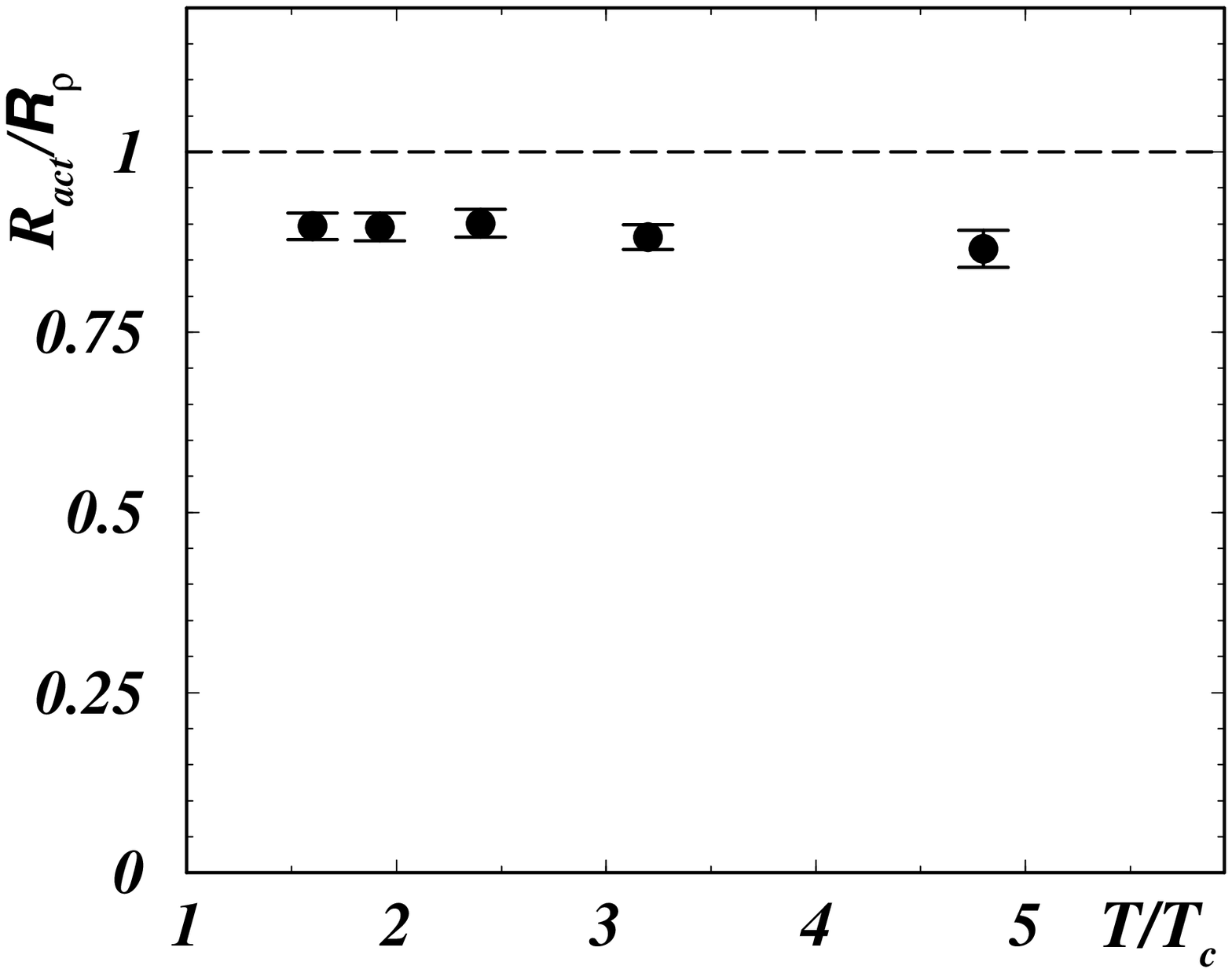,width=6.5cm,height=5.0cm}
{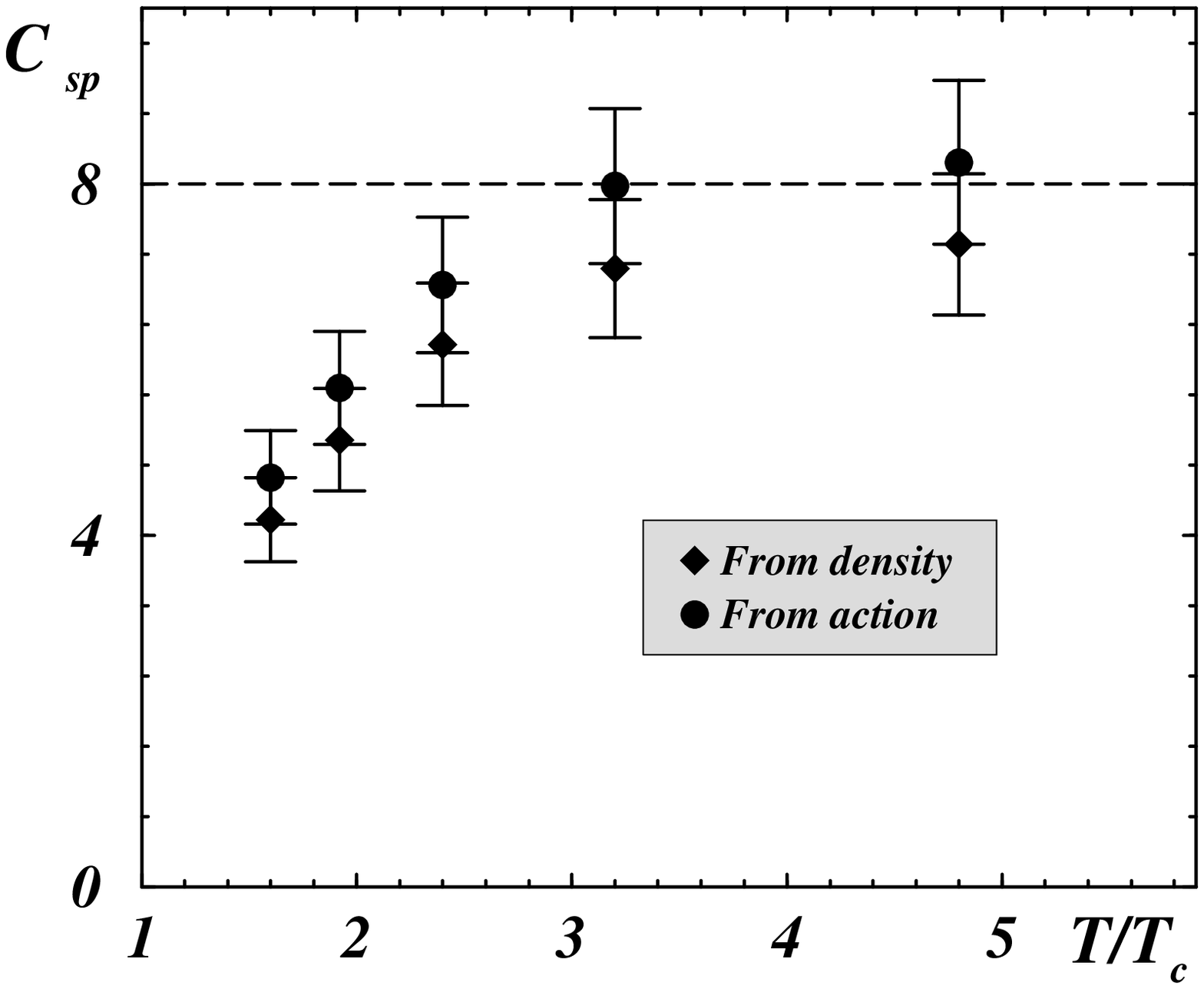,width=6.5cm,height=5.0cm}
{Check of the self-consistency: the ratio of the quantities
  $R$, eq.~\protect\eq{R} obtained from the lattice monopole action and
  density.\label{fig:self1}}
{Check of the dilute Coulomb gas picture: quantities $C_{sp}$,
  eq.~\protect\eq{Rsp}, calculated from the action and
  density.\label{fig:self2}}

A check of the validity of the Coulomb gas picture can be obtained
with the help of the quantity
\begin{equation}
C_{sp}(T) = \frac{\sigma_{sp}(T)}{\lambda_D(T)\, \rho(T)} \equiv
\frac{\sigma_{sp}(T)}{R(T)}\,,
\label{Rsp}
\end{equation}
where $\sigma_{sp}$ is the spatial string tension.  In the abelian
projection approach the spatial string tension should be saturated by
the contributions from the static continuum monopoles. In the dilute
Coulomb gas of mo\-no\-po\-les the string tension is~\cite{Polyakov}:
$\sigma= 8 \sqrt{\rho} \slash g_M $ while the screening length is
given by \eq{lambdaD}. These relations imply that in the dilute
Coulomb gas of continuum monopoles we should get $C_{sp} = 8$.

We use the results for the spatial string tension of
ref.~\cite{SigmaSP} in the high temperature $\SU(2)$ gluodynamics.  It
was found that for the temperatures higher than $T \approx 2 T_c$ the
spatial string tension can be well described by the formula:
$\sigma_{sp}(T) = 0.136(11) \, g^4_{4D} (T) \, T^2$, where $g_{4D}(T)$
is the four-dimensional $\SU(2)$ 2-loop running coupling constant,
$$
g^{-2}_{4D}(T) = \frac{11}{12\, \pi^2} \log
\left(\frac{T}{\Lambda_T}\right) +
\frac{17}{44\, \pi^2} \log \left[2 \log \Bigl(\frac{T}{\Lambda_T}\Bigr)\right],
$$ with the scale parameter $\Lambda_T = 0.076(13) \, T_c$. Taking
also into account the relation between the critical temperature and
the zero-temperature string tension~\cite{Heller}, $T_c = 0.69(2) \,
\sqrt{\sigma}$ we calculate the quantity $C_{sp}$ and plot it in
figure~\ref{fig:self2} as a function of the temperature, $T$. If the
Coulomb picture works then $C_{sp}$ should be close to $8$.  {}From
figure~\ref{fig:self2} we conclude that this is indeed the case at
sufficiently high temperatures, $T \slash T_c \gtrsim 2.5$.

\section{Conclusions}
\label{sec4}

In order to describe the lattice monopole dynamics we have proposed to
consider the lattice monopoles as the defects which are blocked from
continuum. In other words the lattice was suggested to be a measuring
device for the continuum monopoles. As a result we are able to draw
the following conclusions:

\begin{itemize}
\item Using the Monte Carlo results for the density of the squared
  monopole charges and the monopole action we are able to calculate
  the product of the abelian magnetic screening length and the
  monopole density corresponding to the continuum Coulomb gas model.
  The values of this parameter obtained from density and action
  measurements are consistent with each other and --- at sufficiently
  high temperatures --- are consistent with known results for the
  spatial string tension. At temperatures $T \slash T_c \gtrsim 2.5$
  the spatial string tension is dominated by contributions from the
  continuum static monopoles.
\item The \emph{continuum} Coulomb gas model can describe the results
  of the Monte Carlo simulations for the action and density of the
  \emph{lattice} mo\-no\-po\-les. The dependence of these quantities
  on the physical sizes of the lattice mo\-no\-po\-les (the size of
  the cell which is used for the monopole detection) is in a good
  agreement with the analytical predictions.
\item The lattice monopole action is dominated by the mass and the
  Coulomb terms for, respectively, small and large sizes of the {\it
    lattice} monopoles. A relation between the density of the squared
  magnetic charges and the monopole action is established
  (eqs.~(\ref{mon:tree}), (\ref{rho:b}), (\ref{P}) and/or
  eqs.~(\ref{Coul}), (\ref{Mass}), (\ref{CM})). Our analytical
  derivation was done in a gaussian approximation. We have shown that
  the corrections to the leading terms of the action are small in the
  large-$b$ limit. A detailed analysis of the corrections will be
  discussed elsewhere~\cite{in:preparation}.
\end{itemize}

We believe that this method can also be applied to the four
dimensional non-abelian gauge theory. We think that this would allow
to get (at least, a part of) parameters of the dual superconductor
model.

\acknowledgments

We thank E.M.~Ilgenfritz, M.I.~Polikarpov, H.~Reinhardt and
V.I.~Zakharov for interesting discussions. We also grateful to
W.~Bietenholz for bringing to our attention
refs.~\cite{Wolfgang1,Wolfgang2}.  M.N.Ch.\ is supported by the JSPS
Fellowship P01023.

\end{document}